\documentclass[acmsmall,nonacm]{acmart}
\usepackage{tcolorbox}
\usepackage{amsmath}
\usepackage{hyperref}

\AtBeginDocument{%
  }

\begin{document}

\title{ReDef: Do Code Language Models Truly Understand Code Changes for Just-in-Time Software Defect Prediction?}

\author{Doha Nam}
\email{waroad@kaist.ac.kr}
\orcid{0009-0006-4195-2931}
\affiliation{%
  \institution{Korea Advanced Institute of Science and Technology(KAIST)}
  \city{Daejeon}
  \country{South Korea}
}

\author{Taehyoun Kim}
\email{tae_hyoun@kaist.ac.kr}
\orcid{0009-0001-0691-4255}
\affiliation{%
  \institution{Korea Advanced Institute of Science and Technology (KAIST)}
  \city{Daejeon}
  \country{South Korea}
}
\affiliation{%
  \institution{Agency for Defense Development (ADD)}
  \city{Daejeon}
  \country{South Korea}
}

\author{Duksan Ryu}
\email{duksan.ryu@jbnu.ac.kr}
\orcid{0000-0002-9556-0873}
\affiliation{%
  \institution{Jeonbuk National University}
  \city{Jeonju}
  \country{South Korea}
}

\author{Jongmoon Baik}
\email{jbaik@kaist.ac.kr}
\orcid{0000-0002-2546-7665}
\affiliation{%
  \institution{Korea Advanced Institute of Science and Technology(KAIST)}
  \city{Daejeon}
  \country{South Korea}
}

\renewcommand{\shortauthors}{Nam et al.}


\begin{abstract} 
Just-in-Time software defect prediction (JIT-SDP) plays a critical role in prioritizing risky code changes during code review and continuous integration. However, existing datasets often suffer from noisy labels and low precision in identifying bug-inducing commits. To address this, we present \textbf{ReDef} (\textbf{Re}vert-based \textbf{Def}ect dataset), a high-confidence benchmark of function-level modifications curated from 22 large-scale C/C++ projects. Defective cases are anchored by revert commits, while clean cases are validated through post-hoc history checks. Ambiguous instances are conservatively filtered out via a GPT-assisted triage process involving multiple votes and audits. This pipeline yields 3,164 defective and 10,268 clean modifications, offering substantially more reliable labels than prior resources. Beyond dataset construction, we provide a systematic evaluation of how Code Language Models (CLMs)---specifically CodeBERT, CodeT5+, UniXcoder, and Qwen2.5---reason about code modifications. We first investigate which input encodings most effectively expose change information under five different strategies. We then design four counterfactual perturbation strategies (e.g., swapping added/deleted blocks, inverting diff polarity) to serve as diagnostic probes. We posit that if models genuinely capture change semantics, such distortions should lead to a clear decline in predictive performance. Our results show that compact diff-style encodings consistently outperform whole-function formats across all CLMs, supported by rigorous statistical confirmation. However, under counterfactual tests, performance remains effectively stable, revealing that what appears to be robustness in fact reflects a reliance on superficial cues rather than true semantic understanding. These findings indicate that, at least in code-change understanding tasks, current CLMs remain limited in their ability to genuinely comprehend the relational dynamics of code modifications. \end{abstract}




\maketitle

\section{Introduction}
Software defect prediction (SDP) aims to estimate whether a given piece of code is likely to contain defects so that limited engineering resources---testing, review time, and Continuous Integration (CI) capacity---can be prioritized where they matter most~\cite{malhotra2015systematic}. Unlike defect detection or localization, which seeks to identify concrete bug instances, SDP is a risk estimation problem used to triage changes and guide quality assurance activities. Effective SDP reduces downstream failures and lowers maintenance costs by surfacing risky code early enough for developers to take action within their normal workflow---for example, during code review or before a merge in CI.

Classical SDP typically models static snapshots of code or aggregates hand-crafted process metrics---such as churn, complexity, ownership, and past defects---at the file, module, or function level~\cite{wahono2015systematic}. While such predictors can correlate with risk, they fail to capture how the code is evolving at the moment of change. Just-in-Time (JIT) SDP addresses this limitation by reasoning over code modifications at commit time, aligning closely with practical review and CI processes that scrutinize the changed lines~\cite{kamei2012large, kim2008classifying}. However, most existing JIT-SDP datasets rely on SZZ(Śliwerski--Zimmermann--Zeller)-style linkages from bug-fixing commits back to bug-introducing commits~\cite{sliwerski2005changes}. SZZ attributes the introduction of a defect to the last modifier of the lines changed by a later fix---an assumption that is often violated by routine edits such as renaming or refactoring. This approach introduces substantial noise: it can misidentify the true introduction points, propagate commit-level label errors, and mismatch the actual defective lines, ultimately obscuring evaluation quality~\cite{herbold2022problems,lyu2024evaluating}. To mitigate the noise inherent in SZZ-based labeling, we turn to a more explicit and high-confidence signal: revert commits.

Reverts are a routine part of disciplined software development. Developers may revert commits for various reasons---such as regressions, incompatibilities, performance degradations, policy or formatting violations, or simply because another change has superseded the patch~\cite{yan2019characterizing}. A revert creates an explicit, time-stamped link to the earlier change. When the revert message cites a failure or fix, it provides strong evidence that the prior modification was defective. We therefore use revert commits as anchors to harvest high-confidence positive examples, and subsequently filter out cases unrelated to defective modifications.

With a more reliable labeling strategy in place, the next question is how modern models can be effectively leveraged. Existing JIT-SDP research has largely depended on compact engineered features with classical machine learning (ML) and, more recently, deep learning (DL) models. However, these approaches primarily capture surface-level correlations---such as churn, ownership, and developer activity---rather than the true semantics of code changes~\cite{kamei2012large, kamei2016studying, hoang2019deepjit}. In other words, they estimate probabilities from shallow cues instead of reasoning about what the code actually does.

By contrast, recent Code Language Models (CLMs)---ranging from specialized encoder-based architectures to large-scale generative decoders (LLMs)---have demonstrated an unprecedented ability to capture the deep semantic meaning of code, moving beyond mere statistical correlates. This advancement has led to state-of-the-art performance across diverse software engineering tasks, including code search, summarization, repair, and vulnerability detection~\cite{guo2022unixcoder, xia2023automated, ni2023function, hui2024qwen2}. This success raises a fundamental question for JIT-SDP: if CLMs can effectively understand static code snapshots, can they also reason about code modifications (i.e., the causal relationship between the before and after states)? Given that CLMs, regardless of their specific architecture, have consistently shown superior effectiveness in code understanding, we focus our study on evaluating these models as the primary candidates for a systematic evaluation of change-level defect prediction~\cite{zhang2025benchmarking, sun2023automatic, zhang2025revisiting}.

To answer this question, we first construct a high-confidence, change-centric benchmark. This high label precision is essential for a rigorous analysis, as it ensures that any observed model limitations stem from a lack of semantic understanding rather than noise in the ground truth. Next, in Stage 1, we compare alternative ways of presenting modification information to CLMs, asking which input encodings yield the most effective performance.
For this, we fine-tune four widely used CLMs---CodeBERT~\cite{feng2020codebert}, CodeT5+~\cite{wang2023codet5+}, UniXcoder~\cite{guo2022unixcoder}, and Qwen2.5~\cite{hui2024qwen2}---as backbones.
However, strong results under these encodings alone do not guarantee genuine semantic reasoning, as models may still rely on superficial cues. Therefore, in Stage 2, we deliberately perturb the representation of code changes---e.g., by swapping added and deleted code blocks, inverting edit polarity, or injecting spurious markers---to test whether CLMs degrade when true semantic information is distorted.
By probing how models behave under such distortions, our protocol not only measures raw accuracy but also uncovers a blind spot in current CLM research: apparent robustness may simply reflect reliance on shallow cues rather than genuine semantic reasoning about code changes.

\noindent Our key contributions are as follows:
\begin{itemize}
  \item \textbf{High-Confidence JIT-SDP Dataset (ReDef):} We curate ReDef (\textbf{Re}vert-based \textbf{Def}ect dataset), a corpus of single-function changes harvested via revert anchors and refined through GPT-assisted filtering. The dataset is further validated through manual sampling to ensure benchmark reliability, and its 10K+ (3,164 defective, 10,268 clean) scale makes it well-suited for fine-tuning CLMs.
  \item \textbf{Encoding Strategies for CLMs:} We conduct the first systematic comparative study of five encoding strategies for representing code modifications as CLM inputs, providing clear empirical evidence of how input representations affect change-centric defect prediction.
  \item \textbf{Counterfactual Robustness Analysis:} We design counterfactual perturbation tests to probe whether CLMs truly reason about code changes beyond surface cues, revealing limitations of current models and offering guidance for future research.
\end{itemize}

Together, these contributions provide both a reliable benchmark and empirical insights into how CLMs handle code changes, informing future defect prediction research and practice.
The remainder of this paper is organized as follows. Section~2 reviews background and related work. Section~3 describes our dataset construction and two-stage evaluation protocol, encompassing both input encoding strategies and counterfactual perturbation methods. Section~4 outlines the experimental setup, and Section~5 reports results. Section~6 discusses threats to validity, and Section~7 concludes.

\section{Background and Related Work}
This section reviews prior research on software defect prediction and the challenges of dataset reliability. 
Section~2.1 briefly outlines classical \emph{Software Defect Prediction (SDP)}, which focuses on identifying defect-prone modules from static snapshots. 
Section~2.2 covers the evolution of \emph{Just-in-Time software defect prediction (JIT-SDP)}, focusing on the transition from hand-crafted metrics to deep learning and recent Code Language Models (CLMs). 
Finally, Section~2.3 discusses the critical limitations of existing JIT-SDP datasets, specifically the noise and reliability issues inherent in SZZ-based labeling, which motivates our construction of the ReDef benchmark.

\subsection{Artifact-level Software Defect Prediction (SDP)}
Classical SDP focuses on predicting defect-prone artifacts (e.g., files or modules) from static snapshots. Early approaches relied on hand-crafted product metrics (e.g., size, complexity, coupling) and process metrics (e.g., churn, developer ownership, past defects)~\cite{malhotra2015systematic, bell2013limited, huda2017framework} fed into traditional machine learning models like SVMs and random forests~\cite{elish2008predicting, balaram2022prediction, wang2012compressed, khoshgoftaar2002tree}. More recent work has shifted toward learned representations using deep generative models or transformer-based encoders to capture richer features~\cite{wang2018deep, omri2020deep}. While these methods have evolved to consider various practical evaluation settings~\cite{ostrand2007measure, tosun2009practical}, snapshot-based prediction remains limited as it cannot capture the semantics of specific code changes and often blurs defective regions at a coarse granularity. This lack of change-level precision necessitates Just-in-Time (JIT) approaches that provide actionable insights at the time of commit.

\subsection{Just-in-Time Software Defect Prediction (JIT-SDP)}
JIT-SDP shifts the focus from static artifacts to change-time risk estimation, scoring each commit or patch so that reviewers and CI gates can provide immediate feedback. The concept originates from Kim et al., who first classified individual changes as clean or buggy~\cite{kim2008classifying}. Subsequent work engineered compact commit-level features---such as churn, diffusion, prior defect density, developer experience, recency, and message cues---and applied machine learning techniques~\cite{kamei2012large, shihab2012industrial, kamei2016studying}. Later studies incorporated commit messages and patch text, while neural models began encoding diffs directly. For example, Yang et al.\ proposed Deeper, a Deep Belief Network (DBN)-based model that outperformed logistic regression~\cite{yang2015deep}, and Hoang et al.\ introduced DeepJIT, which integrates commit messages and code metrics~\cite{hoang2019deepjit}. CC2Vec further embedded code changes hierarchically to capture structural and sequential information~\cite{hoang2020cc2vec}.

More recently, researchers have applied CLMs to JIT-SDP. Guo et al.\ systematically compared six transformer-based backbones (RoBERTa, CodeBERT, BART, PLBART, GPT2, and CodeGPT) and showed that CLM-based variants consistently outperform traditional baselines such as DeepJIT and CC2Vec~\cite{guo2023study}. Building on this, Abu Talib et al. demonstrated that parameter-efficient fine-tuning (PEFT) techniques such as LoRA can further boost JIT-SDP performance on ApacheJIT~\cite{abu2024parameter}, while Lin et al. introduced CCT5---a CodeT5-initialized model pre-trained on CodeChangeNet---that achieves state-of-the-art results on code-change tasks including JIT-SDP~\cite{lin2023cct5}. Jiang et al.\ extended CodeBERT with a Replaced Message Identification objective, resulting in BiCC-BERT and JIT-BiCC, which improved F1-scores by over 10\%~\cite{jiang2025just}. 
Together, these studies highlight the promise of CLMs for JIT-SDP, appearing to possess an advanced capability to comprehend code modifications. However, most existing research has focused on compact encoder-based or encoder-decoder models, leaving the potential of contemporary large-scale decoder-based models---such as the Qwen~\cite{yang2025qwen3} or Llama families~\cite{grattafiori2024llama}---largely underexplored in this domain. Furthermore, it remains unclear whether these performance gains reflect genuine semantic understanding or a reliance on superficial textual patterns, underscoring the need for a systematic evaluation of model behavior.

\subsection{Labeling Reliability and SZZ Limitations}
A precise and high-confidence dataset is a fundamental prerequisite for evaluating whether Code Language Models truly understand the semantics of code changes. 
Despite the technical progress in neural modeling described in Section~2.2, a critical bottleneck in JIT-SDP remains the reliability of dataset labeling.
Most existing JIT-SDP datasets rely on SZZ-based labeling, which identifies bug-introducing commits by tracing the origins of the code lines modified in bug-fixing commits~\cite{sliwerski2005changes}. However, this approach introduces substantial noise that propagates labeling errors and obscures evaluation quality. Herbold et al.\ showed that SZZ-based labeling is highly error-prone, with only about half of the commits identified as bug-inducing being correct~\cite{herbold2022problems}. Even after additional refinement using a RoBERTa-based issue classifier, Afric et al.\ reported that datasets still contained, on average, 14.3\% mislabeled instances~\cite{afric2023empirical}. Lyu et al.\ conducted a large-scale evaluation on over 76K developer-labeled Linux kernel commits and showed that SZZ variants suffer from a recall degradation of 13.8\%, with 17.47\% of bug-fixing commits being ghost commits~\cite{lyu2024evaluating}.

To mitigate these errors, several SZZ variants have been proposed~\cite{fan2019impact}. 
Early improvements like Annotation Graph SZZ (AG-SZZ) attempted to filter non-semantic changes (e.g., whitespace and comments)~\cite{kim2006automatic}. 
Subsequent variants addressed meta-changes (MA-SZZ)~\cite{da2016framework} and automated refactoring detection (RA-SZZ)~\cite{neto2018impact} to prevent incorrect blame assignment. 
More recent efforts include TC-SZZ, which iteratively traces deeper histories to recover missed commits~\cite{lyu2024evaluating}, and Code Change Tactics (CAT), which disentangles refactoring-propagated changes from actual defect-inducing edits~\cite{niu2025refactoring}. 
Yet, even with these sophisticated refinements, automated labeling still struggles with the inherent ambiguity of commit histories.

Fundamentally, pinpointing the root cause of a defect requires deep domain expertise; without explicit documentation in commit messages, it is notoriously difficult for external researchers---or automated tools---to accurately assign blame to specific code modifications. 
Furthermore, even manual inspection by non-developers remains subjective and error-prone, as identifying a bug's origin from raw diffs alone lacks necessary semantic grounding.
These persistent reliability gaps underscore the need for more conservative and evidence-based labeling strategies that rely on explicit developer feedback rather than speculative history tracking. In this paper, we address this need by introducing the ReDef benchmark, which prioritizes high-confidence signals to provide a more robust environment for assessing the true reasoning capabilities of various Code Language Models, spanning encoder-based models (CodeBERT, UniXcoder, and the encoder of CodeT5+) and a large-scale decoder-based model (Qwen2.5).

\section{Methodology}

\begin{figure}[h]
  \centering
  \includegraphics[width=\linewidth]{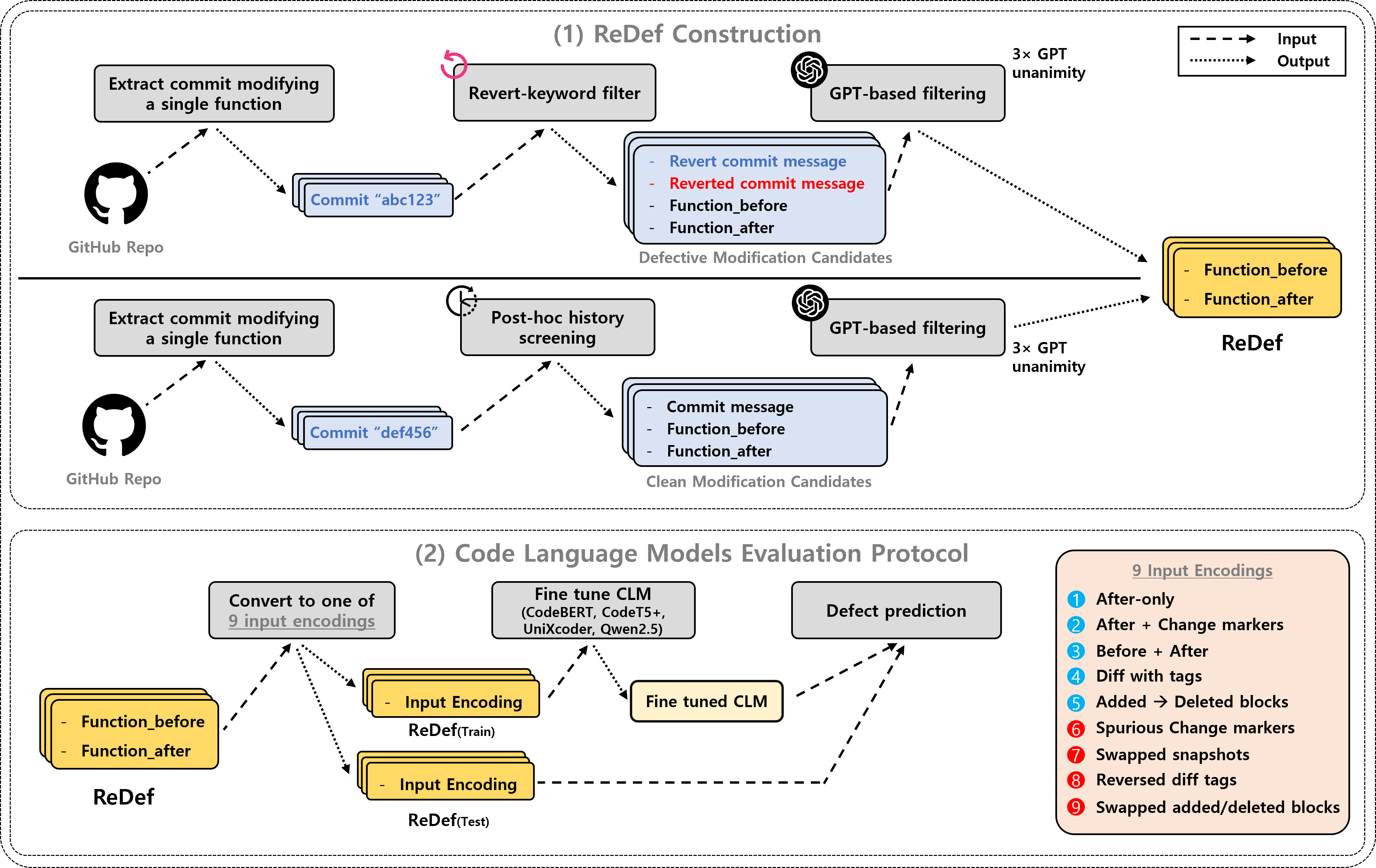}
  \caption{Overall approach of ReDef construction and CLM evaluation protocol.}
  \label{fig:figure1}
\end{figure}

Our methodology comprises two main pillars. (A) ReDef Construction: We curate defective and clean code modifications
from large-scale C/C++ projects, leveraging revert signals, post-hoc history screening, and expert audits to ensure high label reliability.
(B) Code Language Models Evaluation Protocol: We design a two-stage framework to assess how various Code Language Models (CLM) interpret code modifications. Figure~\ref{fig:figure1} summarizes the overall approach.

\subsection{ReDef Construction}
\begin{figure}[h]
  \centering
  \includegraphics[width=1.0\linewidth]{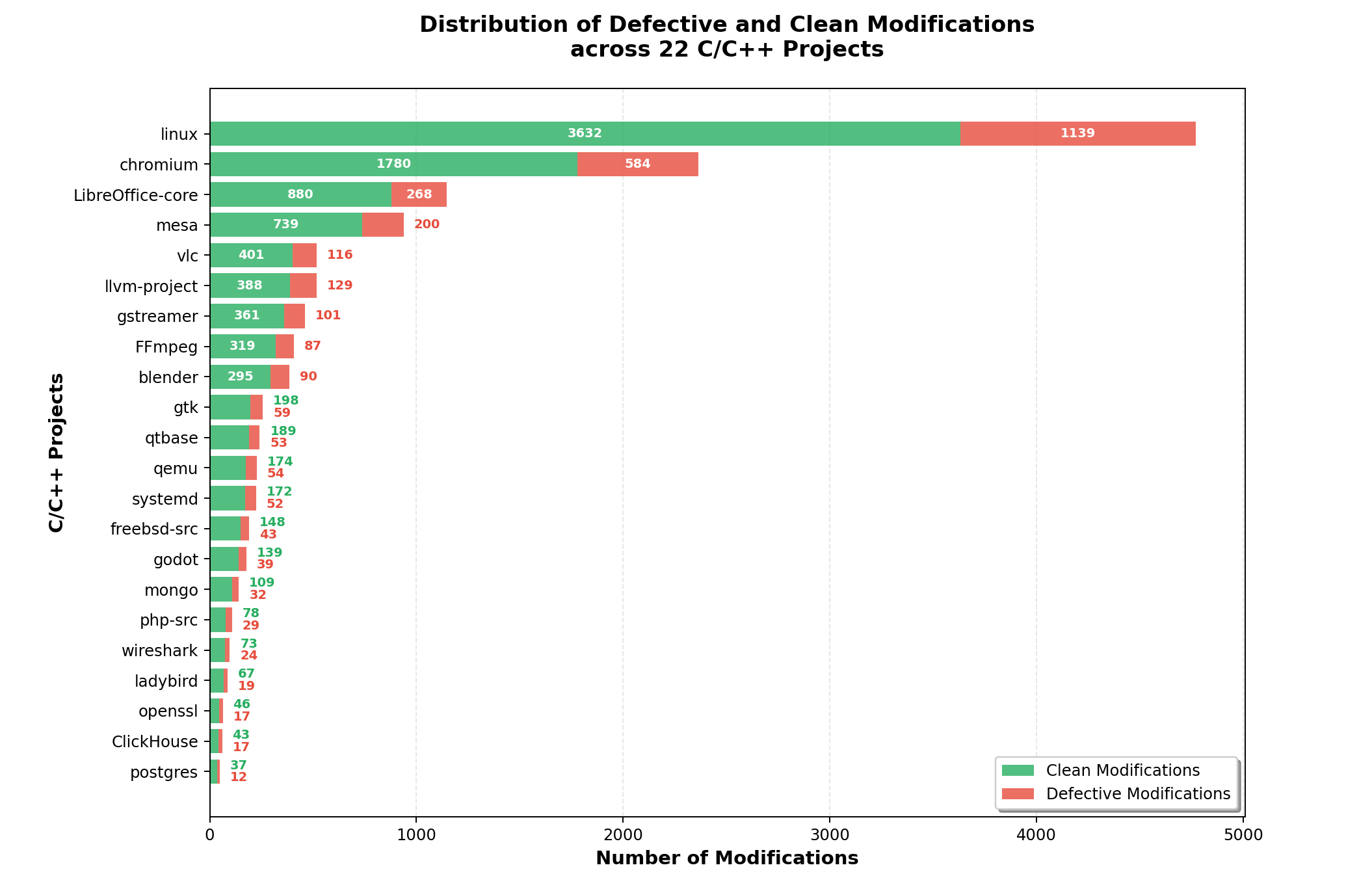}
  \caption{Number of defective and clean modifications per project in the ReDef corpus.}
  \label{fig:figure2}
\end{figure}

We curated ReDef, a high-confidence JIT-SDP corpus from 22 long-lived, high-activity C/C++ repositories. Projects were selected for sustained commit volume, explicit revert workflows, and domain diversity to reduce project-specific bias. The curation yielded 3,164 defective and 10,268 clean modifications. Figure~\ref{fig:figure2} reports per-project counts.

\noindent\textbf{Defective Modification.}
We extracted commits that touched exactly one file and whose modified lines were confined to a single function.
This design ensures reliable causal attribution to a specific function---a choice also adopted in recent datasets such as PrimeVul~\cite{ding2024vulnerability}.
From this single-function pool, we identified defective modifications by collecting revert commits whose messages
(i) contained the keyword ``revert'' and (ii) explicitly referenced the hash of an earlier commit.
Since such reverts typically indicate that the referenced commit introduced a problem,
the targeted commits are highly likely to represent defective modifications.

For each referenced hash, we retrieved its associated commit (the target of the revert)
and collected four artifacts for analysis: the \texttt{reverted commit message},
the \texttt{function\_before} and \texttt{function\_after} from that commit, and the \texttt{revert commit message} itself.
Together these formed an evidence bundle and were submitted to GPT-4o three independent times under a fixed rubric to remove ambiguous or non-defect reverts (e.g., formatting changes, policy reversions, or hash-only messages without rationale). The rubric distinguished each revert as: (A) Failed/incomplete fix (Defective$\rightarrow$Defective), (B) Bug introduction (Clean$\rightarrow$Defective), or (C) Other (non-defect). Categories (A) and (B) were merged into the defective class, while (C) served as a filter to discard non-defect cases.
To ensure high precision, a candidate was retained only if \emph{all three GPT-4o votes unanimously fell in \{A,B\}} (e.g., A/A/A or A/B/B). Any sample with one or more (C) votes (e.g., A/A/C) was conservatively discarded.

Across the 22 subject projects, we identified approximately 0.43M reverts out of 3.37M fix-related commits, indicating that reverts account for roughly 13\% of the total defect pool. This proportion further decreases after applying the single-function constraint and GPT-4o triage. While this design inevitably limits coverage, it reflects a deliberate trade-off to prioritize label precision and mitigate the ``falsely blaming'' issue inherent in traditional SZZ-based heuristics. In contrast to heuristic lineage tracking, revert commits provide explicit causal links between bug fixes and their corresponding bug-inducing changes, and our filtering pipeline removes tangled or ambiguous cases. As a result, ReDef complements existing larger but noisier datasets by offering a controlled, high-confidence benchmark for evaluating the semantic reasoning capabilities of code models.

\noindent\textbf{Clean Modification.}
To assemble the clean stratum, we first identified all defective modifications and then targeted clean modification commits from a similar time period in the same repository,
thereby mitigating both project-specific and period-specific bias.
As with the defective side, we extracted commits that touched exactly one file and whose modified lines were confined to a single function.
For each candidate commit, we inspected up to five subsequent commits that touched the same function.
If any of these later messages contained problematic keywords (e.g., ``revert,'' ``rollback,'' ``fix bug,'' ``regression''),
the candidate was discarded to avoid false-clean labels---since such signals in subsequent edits suggest that the earlier commit may have introduced issues---an idea conceptually similar to SZZ methodology.
Surviving candidates were further required to be modified at least once after their introduction, which excluded dead or unused code.
We then formed an evidence bundle---\texttt{commit message}, \texttt{function\_before}, and \texttt{function\_after}---
and queried GPT-4o three independent times under a dedicated clean-label rubric. The rubric classified each commit as: (A) Bug fix (Defective$\rightarrow$Clean), (B) Improvement/optimization/feature change (Clean$\rightarrow$Clean), or (C) Ambiguous/irrelevant commits. Categories (A) and (B) were retained as clean modifications, while any case with a (C) vote was filtered out. The sample was kept only when all three votes unanimously fell in \{A,B\} (e.g., A/A/A or A/B/B).

Table~\ref{tab:filtering} summarizes the filtering outcomes. The defective pool was reduced from 6,257 to 3,164---about half of revert-anchored commits were discarded, mostly because the reasons for the reverts were not sufficiently specified. Among the retained cases, 1,270 were classified as Defective$\rightarrow$Defective and 1,795 as Clean$\rightarrow$Defective. By contrast, the clean pool was only modestly reduced (11,807 to 10,268), with 6,524 Defective$\rightarrow$Clean and 3,744 Clean$\rightarrow$Clean cases retained, all meeting the unanimous voting criterion. The exact GPT prompts and rubric are available in the provided artifact.

\begin{table}[t]
\centering
\caption{Filtering outcomes for defective (revert-anchored) and clean (history-screened) candidates.}
\label{tab:filtering}
\resizebox{0.7\textwidth}{!}{%
\begin{tabular}{lrr}
\toprule
\textbf{Category} & \textbf{Before GPT Filtering} & \textbf{After GPT Filtering} \\
\midrule
Defective total        & 6,257 & \textbf{3,164} \\
\quad Defective$\rightarrow$Defective & -- & 1,270 \\
\quad Clean$\rightarrow$Defective     & -- & 1,795 \\
\midrule
Clean total            & 11,807 & \textbf{10,268} \\
\quad Defective$\rightarrow$Clean     & -- & 6,524 \\
\quad Clean$\rightarrow$Clean         & -- & 3,744 \\
\bottomrule
\end{tabular}%
}
\end{table}

Clean selection was intentionally conservative: a function had to be revised at least once after the candidate commit, and no problematic keywords could appear in the next five commits touching that function. Unlike prior JIT corpora (e.g., SZZ-derived~\cite{kim2008classifying, kamei2016studying}) or static snapshot datasets (e.g., Devign~\cite{zhou2019devign}, Big-Vul~\cite{fan2020ac}, Defects4J~\cite{just2014defects4j}) that define ``clean'' at a snapshot or immediate post-fix state, our per-function, post-hoc history check aimed to reduce false-clean labels arising from late-emerging faults or follow-up fixes.

\noindent\textbf{Data Partitioning and Statistics.}
After assembling the labels, each repository was split time-respectively (80\%/10\%/10\% into train/validation/test by commit timestamp) to avoid temporal leakage---training on commits occurring after the test set would otherwise expose future information.
Class balance (defective/clean) was preserved within each split, and splitting was performed per repository to ensure that every project was represented across partitions, preventing cases where a project might be entirely absent from validation or test. The per-project partitions were then concatenated to form the final splits.
Table~\ref{tab:dataset_stats} summarizes the scale of the curated modifications. 
On average, the studied functions consist of 94.52 Lines of Code (LOC), with modifications typically affecting 22.82\% of the code. The gradual increase in average LOC from training to test sets reflects the natural growth of project complexity over time, as our chronological split captures the inherent evolution of software projects. Conversely, the slight decrease in the percentage of changed code suggests that modifications become more localized as projects mature and stabilize.

\begin{table}[t]
\centering
\caption{Descriptive statistics of the ReDef dataset.}
\label{tab:dataset_stats}
\resizebox{0.4\textwidth}{!}{%
\begin{tabular}{lrrr}
\toprule
\textbf{Split} & \textbf{Count} & \textbf{Avg. LOC} & \textbf{\% Changed} \\ \midrule
Train          & 10,744         & 92.12             & 23.30\%             \\
Valid          & 1,344          & 102.29            & 22.01\%             \\
Test           & 1,344          & 105.98            & 19.81\%             \\ \midrule
\textbf{Overall} & \textbf{13,432} & \textbf{94.52}    & \textbf{22.82\%}    \\ \bottomrule
\end{tabular}%
}
\end{table}

\begin{figure}[h]
  \centering
  \includegraphics[width=1.0\linewidth]{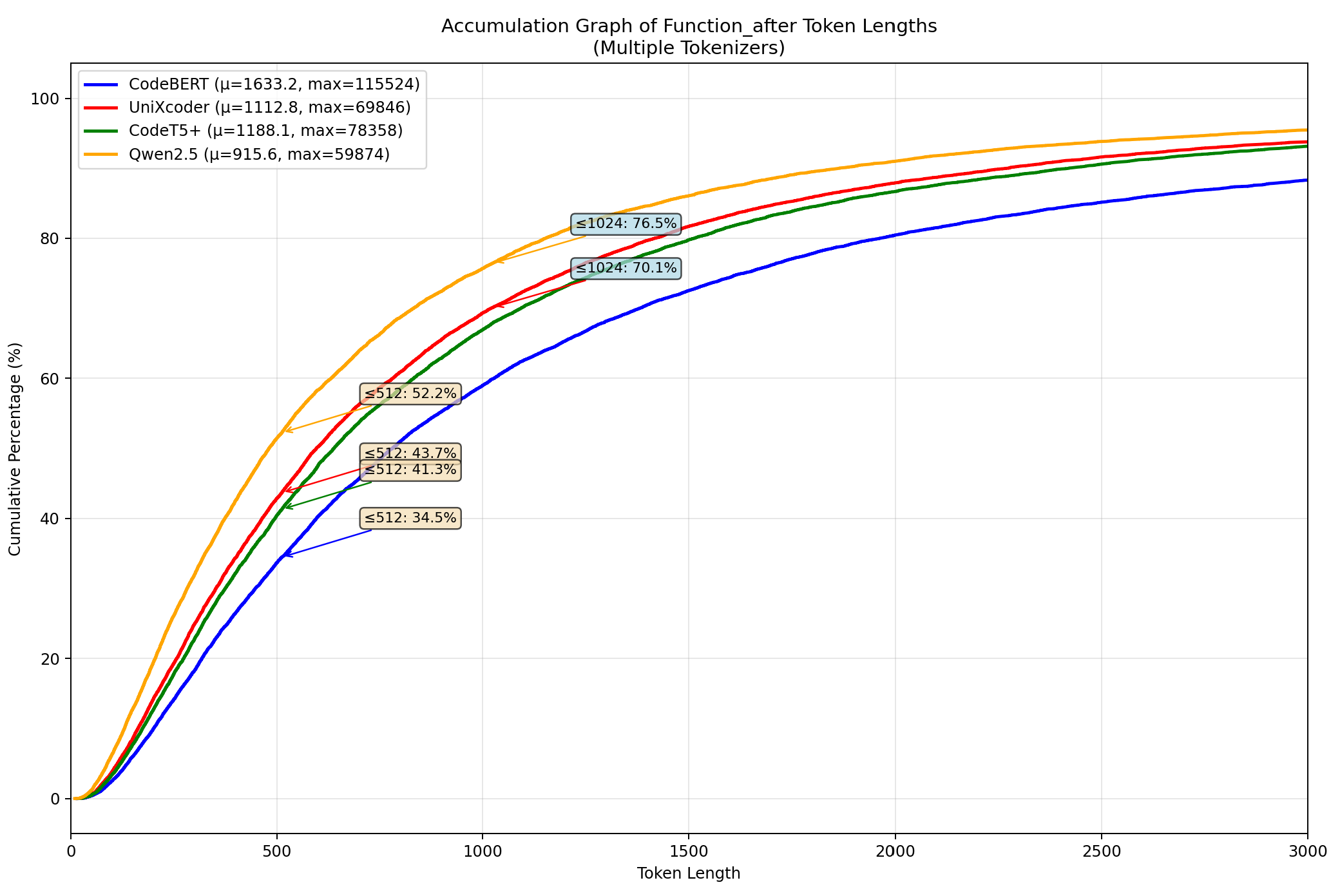}
  \caption{Cumulative distribution of function lengths (tokens) across model tokenizers}
  \label{fig:figure3}
\end{figure}

To characterize input length, we tokenized \texttt{function\_after} using each of the CLMs employed in our experiments (CodeBERT, UniXcoder, CodeT5+, Qwen2.5).
As shown in Figure~\ref{fig:figure3}, the cumulative distribution indicates that although the exact counts vary slightly by model, around 48–65\% of functions exceed the 512-token limit, meaning the
entire function cannot be fed to the model without truncation. With Qwen2.5's 1,024-token limit, roughly 77\% of functions fit in full. This distribution motivates a comparison between encoding strategies that provide only the edited lines versus those that supplying the full function at the risk of truncation. Although the dataset is derived from commit history, the intended inference unit is a function-level modification, which can be evaluated pre-commit (e.g., in an Integrated Development Environment (IDE) or local CI hooks) as well as at review/merge time.
To assess label reliability, two authors independently audited a stratified sample of 100 defective commits under a shared rubric and adjudicated disagreements.

\subsection{Code Language Models Evaluation Protocol}
Our goal is to assess whether Code Language Models (CLMs) can effectively leverage
code-change information for defect prediction. To this end, we designed a two-stage evaluation.
\textbf{Stage~1} compares alternative input encodings that vary in how much context and change signal
are provided. \textbf{Stage~2} introduces counterfactual perturbations that deliberately distort change signals, checking whether models degrade as expected when edit semantics are disrupted.
Together, these stages provide a structured protocol for evaluating whether CLMs rely on genuine semantic cues rather than superficial patterns.

In \textbf{Stage~1}, we compared five alternative input encodings, each tokenized using the model’s native tokenizer before being fed into the CLM encoder.
\begin{enumerate}
    \item \textbf{After-only}: The function\_after, without any explicit indication of where edits occurred. Serves as a classical SDP-style baseline.
    \item \textbf{After+Change markers}: The function\_after, augmented with \texttt{<CHG>} tags on each modified line so the model can directly identify edited regions.
    \item \textbf{Before+After}: Both pre- and post-modification functions concatenated, prefixed with markers (``[Before]: \ldots [After]: \ldots''). To balance length, half of the token budget was allocated to each side, ensuring that the after-code was not entirely truncated by a long before-code.
    \item \textbf{Diff with tags}: Only the changed lines and their local context, with \texttt{<ADD>} and \texttt{<DEL>} tags prepended to additions and deletions respectively, following the representation of DeepJIT~\cite{hoang2019deepjit}.
    \item \textbf{Added$\rightarrow$Deleted blocks}: Inspired by JiT-Fine~\cite{ni2022best}, we supplied added lines first, then deleted lines, each under explicit headers (``[ADDED LINES]: \ldots [DELETED LINES]: \ldots''). Unlike JiT-Fine, commit messages were excluded for fairness.
\end{enumerate}
The first three variants serve as heuristic baselines, representing straightforward ways of exposing code modifications to CLMs without relying on prior proposals.
These five encodings let us quantify the trade-off between context and change signal under realistic token budgets and identify the most effective representation for defect prediction.

In \textbf{Stage~2}, we applied controlled perturbations to test whether models behave consistently when change signals are deliberately altered. Rather than simulating realistic noise, these serve as \emph{diagnostic counterfactuals} to determine whether a model's performance stems from a genuine understanding of code-change semantics (e.g., diff polarity and before/after temporal order) or from an over-reliance on superficial code features. To illustrate, consider a clean bug-fix that changes a risky null-pointer access \texttt{use(p);} to a safe guarded access \texttt{if (p) use(p);}. In the \textbf{Swapped added/deleted blocks} strategy, the representation is transformed as follows:
\begin{center}
\small
\begin{tabular}{|p{0.9\linewidth}|}
\hline
\textbf{Example: Semantic shift in a null-pointer check fix} \\
\textbf{Original (Clean):} \texttt{[DELETED LINES] use(p);} $\rightarrow$ \texttt{[ADDED LINES] if (p) use(p);} \\
$\hookrightarrow$ \emph{A logical fix for a potential crash.} \\
\textbf{Perturbed (Clean):} \texttt{[DELETED LINES] if (p) use(p);} $\rightarrow$ \texttt{[ADDED LINES] use(p);} \\
$\hookrightarrow$ \emph{A logical reintroduction of a bug (regression).} \\
\hline
\end{tabular}
\end{center}
If a model truly understands the semantics of a ``fix,'' its confidence in a ``clean'' label should significantly drop when presented with the perturbed version, which now logically represents a defect. A model whose predictions remain invariant to such reversals is likely relying on the presence of certain keywords (e.g., \texttt{if}, \texttt{NULL}) regardless of their roles in the change, making it unreliable for real-world deployment where distinguishing between a fix and a regression is critical. 

We applied one specific perturbation to each of the four change-aware encodings introduced in Stage~1, excluding the \textbf{After-only} baseline as it lacks explicit change signals to disrupt. We intentionally varied the scope of these perturbations---applying them either across both training and testing or exclusively at test time---to explore two distinct diagnostic scenarios:
\begin{enumerate}
    \item \textbf{Spurious change markers}: In the \textbf{After+Change markers} format, the number of modified lines is first computed, then the same number of \texttt{<CHG>} tags are randomly inserted at arbitrary locations. This injection of meaningless signals occurs during \emph{both training and testing.}
    \item \textbf{Swapped snapshots}: Under the \textbf{Before+After} format, models are trained on original data but \emph{perturbed only at test time} by swapping the positions of \texttt{function\_before} and \texttt{function\_after} with a 50\% probability (e.g., ``[Before]: function\_after [After]: function\_before'').
    \item \textbf{Reversed diff tags}: Under the \textbf{Diff with tags} format, models are trained on original data but \emph{perturbed only at test time} with the \texttt{<ADD>} and \texttt{<DEL>} tags consistently inverted.
    \item \textbf{Swapped added/deleted blocks}: In the \textbf{Added$\rightarrow$Deleted blocks} format, the two sections are randomly swapped with a 50\% probability during \emph{both training and testing}, such that ``[ADDED LINES]'' contains deletions and ``[DELETED LINES]'' contains additions.
\end{enumerate}
Each perturbation targeted the representation of change information while leaving the overall code form largely intact~\cite{ribeiro2018semantically, jia2017adversarial}. These counterfactual variants are designed to reduce model performance: if a CLM truly understands the relationship between code before and after---that is, the semantics of the modification---then disrupting this information should lead to a drop in performance compared to the corresponding Stage~1 results.
Conversely, if performance remains largely unchanged, it suggests that the model has not learned meaningful change semantics and is instead relying on superficial cues from the overall code context.

\section{Experimental Setup}

\subsection{Research Questions}
We try to answer following research questions:

\textbf{RQ1:} \emph{How accurate are the labels in the ReDef JIT-SDP dataset?}

\textbf{RQ2:} \emph{Which input encodings yield the best CLM performance?}

\textbf{RQ3:} \emph{Does CLM performance degrade under counterfactual perturbations of change representations?} 

\subsection{Models and Baselines}
To evaluate how CLMs capture the semantics of code changes, we employed four representative models across distinct architectures. We included established encoder-based models---CodeBERT, CodeT5+, and UniXcoder---which have consistently demonstrated strong performance in JIT-SDP tasks~\cite{jiang2025just, lin2023cct5}. Additionally, we incorporated Qwen2.5, a state-of-the-art decoder-based model. To the best of our knowledge, this study represents the first evaluation of Qwen2.5 within the JIT-SDP domain, despite its reported success in other code-related tasks.

Regarding input constraints, CodeBERT and CodeT5+ were evaluated using their default maximum limit of 512 tokens. In contrast, UniXcoder and Qwen2.5 were tested at both 512- and 1,024-token scales. For Qwen2.5 specifically, we extended the total token budget by 100 tokens (setting the limits to 612 and 1,124) to accommodate the necessary system prompt and role definitions without truncating the actual code content, thereby ensuring a fair comparison of the effective code context.

\paragraph{Subject Models.}
\begin{enumerate}
    \item \textbf{CodeBERT}~\cite{feng2020codebert}
    We included CodeBERT, a 125M parameter RoBERTa-based encoder pre-trained for code understanding. Despite being one of the earlier code language models, it has consistently shown strong and robust performance in defect prediction and classification tasks, making it an essential baseline for fair comparison~\cite{xiao2023empirical, dahou2025semantic}.

    \item \textbf{CodeT5+}~\cite{wang2023codet5+}
    We used the 220M encoder-only variant, as detection tasks do not require generation capabilities. CodeT5+ has achieved state-of-the-art results across a wide range of software engineering tasks~\cite{niu2023empirical, gao2024learning}, making it a competitive and modern baseline.

    \item \textbf{UniXcoder}~\cite{guo2022unixcoder}
    We used the base-nine version (220M parameters, released in 2023), which has demonstrated strong performance across diverse SE tasks~\cite{wang2024line, ding2024vulnerability}.
    
    \item \textbf{Qwen2.5}~\cite{hui2024qwen2}
    We included Qwen2.5-7B-Instruct, representing the current state-of-the-art in open-source decoder-based models. We selected the 7B variant as the standard representative of large-scale models to evaluate its reasoning capabilities over code modifications. To further distinguish between the model's inherent zero-shot reasoning and its task-specific adaptation, we evaluated Qwen2.5 in two settings: an inference-only (zero-shot) baseline and a fine-tuned version.
\end{enumerate}

In addition to these CLMs, we implemented input encodings inspired by \textsc{DeepJIT} (diff with \texttt{<ADD>}/\texttt{<DEL>} tags) and \textsc{JIT-Fine} (added$\rightarrow$deleted block format). These were not used as standalone baselines, but rather as representation strategies integrated into our Stage~1 evaluation protocol.

\subsection{Evaluation Metrics}
We evaluate model performance using five standard classification metrics: Accuracy, Precision, Recall, F1-score(F1), and PR-AUC.
Among these, F1 is adopted as the primary metric, as it balances precision and recall and is widely used for imbalanced datasets such as JIT-SDP~\cite{hoang2019deepjit, kamei2012large, hoang2020cc2vec}.

\begin{equation}
\text{Accuracy} = \frac{TP + TN}{TP + TN + FP + FN}
\end{equation}
Accuracy measures the overall proportion of correctly classified modifications, but can be misleading under severe class imbalance.

\begin{equation}
\text{Precision} = \frac{TP}{TP + FP}, \quad
\text{Recall} = \frac{TP}{TP + FN}
\end{equation}
Precision quantifies the fraction of predicted defective modifications that are truly defective, while Recall measures the fraction of actual defective modifications that are successfully identified.

\begin{equation}
\text{F1} = \frac{2 \cdot \text{Precision} \cdot \text{Recall}}{\text{Precision} + \text{Recall}}
\end{equation}
F1 is the harmonic mean of precision and recall, and is widely regarded as the most informative single measure in imbalanced classification settings.

\begin{equation}
\text{PR-AUC} = \int_{0}^{1} \text{Precision}(r) \, dr
\end{equation}
PR-AUC is the area under the Precision–Recall curve, which provides a threshold-independent view of the trade-off between precision and recall across all operating points.
Note that for the decoder-based model (Qwen2.5), we primarily report the four discrete classification metrics, as calculating threshold-independent measures like PR-AUC is less straightforward due to its generative inference nature compared to encoder-based models with dedicated classification heads.

\subsection{Implementation Details}
To ensure the reproducibility and statistical significance of our results, we fine-tuned each CLM--encoding configuration ten times using different random seeds (12345--12354) and report the mean and standard deviation. All experiments were conducted using NVIDIA Tesla V100 (32GB) and RTX 3080 Ti (12GB) GPUs.

\noindent\textbf{Encoder-based Training.}
CodeBERT, CodeT5+, and UniXcoder were fine-tuned for 6 epochs with a learning rate of $1\mathrm{e}{-5}$, a batch size of 8, and the AdamW optimizer with decoupled weight decay~\cite{loshchilov2017decoupled}. We employed a linear learning-rate scheduler with a 10\% warm-up ratio and utilized gradient accumulation to stabilize training. To address class imbalance, we applied a cost-sensitive weighting scheme, assigning the positive class a weight of $N_{\text{neg}}/N_{\text{pos}}$~\cite{king2001logistic}. The best-performing model across the 6 epochs was selected based on validation F1 for final testing.

\noindent\textbf{Decoder-based Training.}
For Qwen2.5-7B-Instruct, we utilized the Unsloth~\cite{unsloth} framework to perform memory-efficient Supervised Fine-Tuning (SFT) with 4-bit quantization. Given the large parameter scale, we applied Parameter-Efficient Fine-Tuning (PEFT) via LoRA ($r=16, \alpha=16$) to the query, key, value, and projection layers. The model was trained for a single epoch with a learning rate of $2\mathrm{e}{-4}$, an effective batch size of 16, and the 8-bit AdamW optimizer. 
To ensure optimal convergence within a single epoch, we performed intra-epoch validation at 10 equally spaced intervals and selected the checkpoint with the highest validation F1. Unlike the encoder models, class imbalance was addressed via oversampling the defective class to match the size of the clean class~\cite{buda2018systematic}. The same prompt templates were utilized for both fine-tuning and zero-shot evaluations to ensure consistency, all of which are available in our replication package.

\noindent\textbf{Statistical Analysis.}
For RQ2, we conducted a two-way repeated-measures ANOVA with within-subject factors \textit{model} (6) and \textit{encoding} (5), treating the 10 seeds as subjects. We report partial $\eta^2$ and Cohen's $f$ as effect sizes~\cite{lakens2013calculating}. Post-hoc comparisons used paired $t$-tests with Holm step-down correction~\cite{holm1979simple} applied per family: Encoding main-effect pairs on subject-level marginal means (collapsed over models), and all encoding pairs within each model (simple effects).

For RQ3, each perturbation was paired with its corresponding original encoding and evaluated across the same 10 seeds. We tested the directional hypothesis that perturbations degrade performance using one-sided paired $t$-tests ($H_1: F1_{\text{orig}} > F1_{\text{pert}}$), complemented by Wilcoxon signed-rank tests~\cite{wilcoxon1945individual}. We also report paired effect size as Cohen's $d_z$ and non-parametric percentile 95\% bootstrap confidence intervals (CIs) for the mean difference $\Delta = F1_{\text{orig}} - F1_{\text{pert}}$ using 10{,}000 resamples~\cite{tibshirani1993introduction}. $p$-values were adjusted for multiple comparisons using Holm's method within each set of tests (e.g., across the four perturbations).
Following common conventions, we consider results statistically significant at $p<0.05$. Effect sizes are interpreted using standard thresholds: partial $\eta^2 \approx 0.01$ (small), $0.06$ (medium), $0.14$ (large); Cohen's $d_z \approx 0.2$ (small), $0.5$ (medium), $0.8$ (large)~\cite{cohen2013statistical}.

\section{Experimental results}
\noindent\textbf{RQ1:} \emph{How accurate are the labels in the ReDef JIT-SDP dataset?}

We manually audited 100 randomly sampled defective functions. Following recent practices in vulnerability dataset curation~\cite{chen2023diversevul, ding2024vulnerability}, two authors independently inspected the commit evidence---\texttt{reverted commit message}, \texttt{function\_before}, \texttt{function\_after}, and \texttt{revert commit message}---and labeled each case as \emph{Correct}, \emph{Ambiguous}, or \emph{Wrong}. After adjudication, 92 cases were classified as Correct, 5 as Ambiguous, and 3 as Wrong, yielding a positive-label precision of 0.92 as shown in Table~\ref{tab:rq1-audit}. Compared to prior reports of 14–50\% noise in SDP datasets~\cite{herbold2022problems, afric2023empirical}, our 8\% mislabel rate is substantially lower.

\begin{table}[t]
\centering
\caption{Outcomes of the manual audit for 100 randomly sampled defective labels.}
\resizebox{0.5\textwidth}{!}{%
\begin{tabular}{lrrrr}
\toprule
 & \textbf{Correct} & \textbf{Ambiguous} & \textbf{Wrong} \\
\midrule
Reviewer~1   & 94 & 5  & 1 \\
Reviewer~2   & 83 & 14 & 3 \\
Adjudicated  & \textbf{92} & 5 & 3 \\
\bottomrule
\end{tabular}
}
\label{tab:rq1-audit}
\end{table}

Ambiguities initially accounted for ~30\% of cases, typically arising when revert messages
cited only a hash without any rationale, forcing annotators to infer defect-relatedness. We revised
the rubric to abstain unless defect links were explicit, which reduced ambiguous cases to 5\%
in the final adjudicated set. The three Wrong cases followed non-defect patterns (e.g., mistaken commit ordering, temporary mitigations, or structural refactoring).

Overall, these results suggest that our revert-anchored, history-screened pipeline yields comparatively reliable defect labels.

\begin{tcolorbox}[colback=gray!10, colframe=gray!40, title=Summary of RQ1]
A manual audit of 100 defective functions found that 92\% were correctly labeled, with only 8\% identified as ambiguous or wrong. This demonstrates our revert-anchored, history-screened pipeline provides substantially more reliable labels than prior SDP datasets.
\end{tcolorbox}

\noindent\textbf{RQ2:} \emph{Which input encodings yield the best CLM performance?}

Table~\ref{tab:rq2_full} summarizes the performance of all CLM–encoding combinations. Overall, \textbf{Diff w/ tags} and \textbf{Added$\rightarrow$Deleted} consistently achieved the highest F1 across most models, while \textbf{After+Markers} and \textbf{Before+After} often performed worst, particularly in encoder-based models. Notably, we observed a recurring discrepancy where encodings with lower F1 often rank higher in accuracy. This discrepancy reflects the class imbalance in ReDef (3:1 clean-to-defective ratio), where accuracy is often inflated by a bias toward the majority clean class despite poor defect detection performance.

Beyond these task-specific adaptation results, Qwen2.5’s zero-shot performance exhibits a universal decline across all metrics. The \textbf{After-only} zero-shot configuration fails almost entirely (F1 $<$ 0.1), underscoring that in zero-shot settings, providing explicit change information is crucial for the model to establish any predictive utility. To compare the fine-tuned models more rigorously beyond such raw patterns, we applied statistical tests as follows.

\begin{table*}[t]
\centering
\caption{Stage~1 performance analysis of different encodings across models (mean $\pm$ std over 10 runs).}
\label{tab:rq2_full}
\resizebox{\textwidth}{!}{%
\begin{tabular}{llccccc}
\toprule
Model & Encoding & F1 & PR-AUC$^a$ & Accuracy & Precision & Recall \\
\midrule
CodeBERT & After-only     & $0.3731 \pm 0.0191$ & $0.3086 \pm 0.0160$ & $0.5751 \pm 0.0536$ & $0.2901 \pm 0.0201$ & $0.5401 \pm 0.0999$ \\
         & After+Markers  & $0.3524 \pm 0.0238$ & $0.3107 \pm 0.0202$ & $0.6152 \pm 0.0577$ & $0.2968 \pm 0.0175$ & $0.4489 \pm 0.1046$ \\
         & Before+After   & $0.3521 \pm 0.0386$ & $0.3161 \pm 0.0101$ & \underline{$0.6161 \pm 0.0553$} & $0.2999 \pm 0.0197$ & $0.4536 \pm 0.1236$ \\
         & Diff w/ tags   & $0.3953 \pm 0.0101$ & \underline{$0.3349 \pm 0.0076$} & $0.5987 \pm 0.0346$ & $0.3082 \pm 0.0131$ & $0.5571 \pm 0.0602$ \\
         & \underline{\textbf{Added$\rightarrow$Deleted}} & \underline{\textbf{$0.3971 \pm 0.0114$}} & $0.3240 \pm 0.0106$ & $0.5988 \pm 0.0350$ & \underline{$0.3093 \pm 0.0156$} & \underline{$0.5609 \pm 0.0606$} \\
\midrule
CodeT5+  & After-only     & $0.3630 \pm 0.0173$ & \underline{$0.3144 \pm 0.0126$} & $0.6109 \pm 0.0352$ & $0.2977 \pm 0.0150$ & $0.4719 \pm 0.0630$ \\
         & After+Markers  & $0.3381 \pm 0.0279$ & $0.3032 \pm 0.0123$ & $0.6333 \pm 0.0624$ & $0.3010 \pm 0.0191$ & $0.4050 \pm 0.1261$ \\
         & Before+After   & $0.3119 \pm 0.0370$ & $0.3103 \pm 0.0073$ & \underline{$0.6565 \pm 0.0429$} & $0.3023 \pm 0.0166$ & $0.3391 \pm 0.1038$ \\
         & Diff w/ tags   & $0.3764 \pm 0.0223$ & $0.3067 \pm 0.0148$ & $0.6073 \pm 0.0332$ & $0.3028 \pm 0.0187$ & \underline{$0.5050 \pm 0.0733$} \\
         & \underline{\textbf{Added$\rightarrow$Deleted}} & \underline{\textbf{$0.3771 \pm 0.0196$}} & $0.3128 \pm 0.0095$ & $0.6169 \pm 0.0462$ & \underline{$0.3103 \pm 0.0174$} & $0.4965 \pm 0.0948$ \\
\midrule
UniXcoder-512 & After-only     & $0.3495 \pm 0.0308$ & $0.2915 \pm 0.0123$ & $0.5879 \pm 0.0642$ & $0.2833 \pm 0.0138$ & $0.4785 \pm 0.1290$ \\
         & After+Markers  & $0.3210 \pm 0.0327$ & $0.3025 \pm 0.0112$ & \underline{$0.6295 \pm 0.0437$} & $0.2867 \pm 0.0113$ & $0.3789 \pm 0.1013$ \\
         & Before+After   & $0.3196 \pm 0.0528$ & $0.2873 \pm 0.0137$ & $0.5823 \pm 0.0913$ & $0.2691 \pm 0.0055$ & $0.4429 \pm 0.2057$ \\
         & \underline{\textbf{Diff w/ tags}}   & \underline{\textbf{$0.3757 \pm 0.0281$}} & \underline{$0.3197 \pm 0.0147$} & $0.5969 \pm 0.0781$ & \underline{$0.3042 \pm 0.0198$} & \underline{$0.5249 \pm 0.1552$} \\
         & Added$\rightarrow$Deleted & $0.3716 \pm 0.0449$ & $0.3142 \pm 0.0141$ & $0.5984 \pm 0.0725$ & $0.3019 \pm 0.0176$ & $0.5196 \pm 0.1649$ \\
\midrule
UniXcoder-1,024& After-only     & $0.3018 \pm 0.0469$ & $0.2854 \pm 0.0164$ & $0.6211 \pm 0.0499$ & $0.2722 \pm 0.0112$ & $0.3590 \pm 0.1088$ \\
         & After+Markers  & $0.2939 \pm 0.0504$ & $0.3009 \pm 0.0145$ & \underline{$0.6517 \pm 0.0495$} & $0.2880 \pm 0.0103$ & $0.3205 \pm 0.1221$ \\
         & Before+After   & $0.3123 \pm 0.0627$ & $0.2912 \pm 0.0123$ & $0.6027 \pm 0.0841$ & $0.2757 \pm 0.0143$ & $0.4129 \pm 0.2040$ \\
         & \underline{\textbf{Diff w/ tags}}   & \underline{\textbf{$0.3399 \pm 0.0556$}} & \underline{$0.3129 \pm 0.0160$} & $0.6189 \pm 0.1020$ & \underline{$0.3065 \pm 0.0251$} & $0.4451 \pm 0.2209$ \\
         & Added$\rightarrow$Deleted & $0.3352 \pm 0.0684$ & $0.3024 \pm 0.0163$ & $0.6050 \pm 0.0930$ & $0.2911 \pm 0.0173$ & \underline{$0.4580 \pm 0.2313$} \\
\midrule
Qwen2.5-512 & After-only     & $0.3389 \pm 0.0171$ & --- & $0.4957 \pm 0.0450$ & $0.2463 \pm 0.0126$ & $0.5495 \pm 0.0730$ \\
         & After+Markers  & $0.3424 \pm 0.0109$ & --- & $0.4560 \pm 0.0363$ & $0.2400 \pm 0.0090$ & \underline{$0.6009 \pm 0.0485$} \\
         & Before+After   & $0.3382 \pm 0.0130$ & --- & $0.4680 \pm 0.0403$ & $0.2401 \pm 0.0116$ & $0.5767 \pm 0.0528$ \\
         & \underline{\textbf{Diff w/ tags}} & \underline{\textbf{$0.3536 \pm 0.0101$}} & --- & \underline{$0.5115 \pm 0.0385$} & \underline{$0.2579 \pm 0.0114$} & $0.5666 \pm 0.0457$ \\
         & Added$\rightarrow$Deleted & $0.3498 \pm 0.0139$ & --- & $0.4973 \pm 0.0468$ & $0.2528 \pm 0.0139$ & $0.5732 \pm 0.0574$ \\
\midrule
Qwen2.5-1,024 & After-only     & $0.3330 \pm 0.0211$ & --- & $0.4989 \pm 0.0666$ & $0.2448 \pm 0.0107$ & $0.5356 \pm 0.1132$ \\
         & After+Markers  & $0.3407 \pm 0.0135$ & --- & \underline{$0.5081 \pm 0.0365$} & $0.2498 \pm 0.0081$ & $0.5404 \pm 0.0620$ \\
         & Before+After& $0.3442 \pm 0.0155$ & --- & $0.4951 \pm 0.0465$ & $0.2492 \pm 0.0150$ & $0.5615 \pm 0.0511$ \\
         & \underline{\textbf{Diff w/ tags}}   & \underline{\textbf{$0.3505 \pm 0.0138$}} & --- & $0.4945 \pm 0.0526$ & \underline{$0.2529 \pm 0.0145$} & \underline{$0.5782 \pm 0.0611$} \\
         & Added$\rightarrow$Deleted & $0.3396 \pm 0.0179$ & --- & $0.4963 \pm 0.0307$ & $0.2462 \pm 0.0091$ & $0.5511 \pm 0.0662$ \\
\midrule
Qwen2.5-512 & After-only     & $0.0851 \pm 0.0105$ & --- &  \underline{$0.7360 \pm 0.0035$} & $0.2331 \pm 0.0259$ & $0.0521 \pm 0.0067$ \\
(zero-shot) & After+Markers  & $0.2823 \pm 0.0151$ & --- & $0.6365 \pm 0.0064$ & \underline{$0.2641 \pm 0.0130$} & $0.3032 \pm 0.0182$ \\
        & Before+After   & $0.2960 \pm 0.0075$ & --- & $0.6111 \pm 0.0065$ & $0.2583 \pm 0.0071$ & $0.3467 \pm 0.0097$ \\
        & Diff w/ tags   & $0.3274 \pm 0.0058$ & --- & $0.5519 \pm 0.0047$ & $0.2534 \pm 0.0044$ & $0.4625 \pm 0.0097$ \\
        & \underline{\textbf{Added$\rightarrow$Deleted}} & \underline{\textbf{$0.3415 \pm 0.0034$}} & --- & $0.4473 \pm 0.0027$ & $0.2375 \pm 0.0021$ & \underline{$0.6076 \pm 0.0084$} \\
\midrule
Qwen2.5-1,024 & After-only     & $0.0280 \pm 0.0048$ & --- & \underline{$0.7567 \pm 0.0012$} & $0.2421 \pm 0.0333$ & $0.0148 \pm 0.0026$ \\
(zero-shot) & After+Markers  & $0.2823 \pm 0.0158$ & --- & $0.6336 \pm 0.0077$ & \underline{$0.2623 \pm 0.0141$} & $0.3057 \pm 0.0182$ \\
        & Before+After   & $0.2521 \pm 0.0106$ & --- & $0.6263 \pm 0.0043$ & $0.2387 \pm 0.0088$ & $0.2672 \pm 0.0133$ \\
        & Diff w/ tags   & $0.3274 \pm 0.0055$ & --- & $0.5500 \pm 0.0059$ & $0.2529 \pm 0.0048$ & $0.4643 \pm 0.0068$ \\
        & \underline{\textbf{Added$\rightarrow$Deleted}} & \underline{\textbf{$0.3440 \pm 0.0026$}} & --- & $0.4458 \pm 0.0032$ & $0.2386 \pm 0.0019$ & \underline{$0.6161 \pm 0.0049$} \\
\bottomrule
\multicolumn{7}{l}{$^a$PR-AUC is omitted for Qwen2.5 due to its generative output nature.}
\end{tabular}
}
\end{table*}

We conducted a two-way repeated-measures ANOVA with \textit{model} (6 fine-tuned CLMs) and \textit{encoding} (5 variants) as within-subject factors (Table~\ref{tab:anova-results}). Results showed significant main effects of both model ($F(5,45)=9.32, p<.001, \eta^2_p=0.51, f=1.02$) and encoding ($F(4,36)=15.11, p<.001, \eta^2_p=0.63, f=1.30$), both indicating large effect sizes. Post-hoc pairwise tests with Holm correction on the marginal means (collapsed over models) confirmed that \textbf{Diff w/ tags} and \textbf{Added$\rightarrow$Deleted} significantly outperformed all other encodings (all $p_{\text{Holm}} < .01$), while performing comparably to each other.

\begin{table}[t]
\centering
\caption{Two-way repeated-measures ANOVA results with effect sizes.}
\resizebox{0.6\textwidth}{!}{%
\begin{tabular}{lcccccc}
\toprule
Effect & $F$ & df\_num & df\_den & $p$ & $\eta^2_p$ & $f$ \\
\midrule
Model          & 9.32  & 5  & 45  & $<.001$ & 0.51 & 1.02 \\
Encoding       & 15.11 & 4  & 36  & $<.001$ & 0.63 & 1.30 \\
Model$\times$Encoding & 1.91  & 20 & 180 & .014    & 0.17 & 0.46 \\
\bottomrule
\end{tabular}
}
\label{tab:anova-results}
\end{table}

Furthermore, a significant interaction effect was found ($F(20,180)=1.91, p=.014, \eta^2_p=0.17, f=0.46$), suggesting that the impact of encoding strategies varies across different model architectures. Simple effect analysis revealed that for CodeBERT and CodeT5+, change-focused encodings (\textbf{Diff w/ tags}, \textbf{Added$\rightarrow$Deleted}) consistently demonstrated a statistically significant performance advantage ($p_{\text{Holm}} < .05$). In contrast, for UniXcoder and Qwen2.5 models, the pairwise differences did not reach statistical significance. Nevertheless, a consistent numerical advantage of change-focused encodings was observed across \textit{all} models, with \textbf{Diff w/ tags} or \textbf{Added$\rightarrow$Deleted} consistently achieving the highest F1 (Table~\ref{tab:rq2_full}).

This consistent advantage can be explained by two primary factors: truncation risk and distracting context. First, as shown in Figure~\ref{fig:figure3}, only about 30–50\% of functions fit within the 512-token limit. Consequently, whole-function encodings (\textbf{After-only}, \textbf{After+Markers}, \textbf{Before+After}) often truncate important edits, whereas compact diff-style formats preserve the critical change signal within the token budget. Second, supplying the entire function introduces excessive, mostly unchanged context that distracts the model from the actual modification. The weakness of \textbf{Before+After} illustrates both issues: two full functions must fit under the same budget, simultaneously increasing truncation risk and noise. Supporting this distracting context hypothesis, both UniXcoder and Qwen2.5 generally performed better under a 512-token limit than at 1,024 tokens. Therefore, the observed superiority of diff-style encodings is primarily a practical one; it reflects their efficiency in preserving critical signals within finite context windows rather than an inherent semantic advantage over whole-function views.

Interestingly, we observed an inverse correlation between model scale and performance. Despite having the largest parameter scale, Qwen2.5 exhibited the lowest overall performance. This trend extends to encoder-based models, where the model with the fewest parameters---CodeBERT---achieved the highest F1. Given that ReDef consists of difficult samples---modifications that professional developers initially committed and subsequently reverted---the signals for defects are likely extremely subtle. In such high-difficulty scenarios, larger models may struggle to discern meaningful features, while smaller models appear more adept at capturing simpler but consistent patterns.

While the absolute performance remains modest (F1 scores below 0.4), such results are common in classification tasks involving severe class imbalance and subtle signals, as characterized by the ReDef dataset~\cite{guo2023study, ni2023function}. Nevertheless, as all models perform substantially above random baselines, they are evidently capturing discriminative patterns. This observation sets the stage for RQ3, where we investigate whether these learned patterns originate from a genuine understanding of code-change semantics or a reliance on shallow, invariant surface cues.

Finally, in our setting which primarily targets code understanding, CodeBERT again emerged as the strongest CLM, in line with recent studies showing its robustness on classification and comprehension tasks~\cite{xiao2023empirical, dahou2025semantic}.

\begin{tcolorbox}[colback=gray!10, colframe=gray!40, title=Summary of RQ2] 
\textbf{Diff-style encodings} (\textbf{Diff w/ tags}, \textbf{Added$\rightarrow$Deleted}) achieved the best performance by emphasizing localized modifications and minimizing truncation risks. \textbf{CodeBERT} showed the strongest performance overall, while performance tended to decrease as model complexity increased (e.g., Qwen2.5) for the subtle defect signals in ReDef. Finally, despite the zero-shot capabilities of decoder models like Qwen2.5, task-specific fine-tuning remained indispensable for establishing reliable predictive utility. \end{tcolorbox}

\noindent\textbf{RQ3:} \emph{Does CLM performance degrade under counterfactual perturbations of change representations?}

To address RQ3, we conducted counterfactual perturbations using CodeBERT, the best-performing encoder-based model, and Qwen2.5-512, the strongest decoder-based variant. As shown in Table~\ref{tab:rq3_full}, across the four perturbations---\textbf{Spurious change markers}, \textbf{Swapped snapshots}, \textbf{Reversed diff tags}, and \textbf{Swapped blocks}---performance for both CodeBERT and Qwen2.5-512 remained effectively stable. Even when the logical direction or structural alignment of the modification was disrupted, F1 differences stayed within a negligible range ($\Delta \leq 0.0126$). This stability is counterintuitive: if these models were truly leveraging semantic signals from code changes, such distortions should have produced clear drops in performance metrics. Instead, the negligible differences suggest that the models are not semantically sensitive to modification content, but rather exploit surface-level regularities (e.g., token distributions, positional patterns) that persist under perturbation. In other words, the robustness observed here reflects insensitivity rather than genuine understanding.

The lack of response in the Swapped snapshots and Swapped blocks conditions ($p = 1.00$ after Holm correction for both models) provides compelling evidence of semantic blindness. Logically, reversing the states or blocks represents a functional opposite, yet both models were impervious to these changes. One might attribute this stability to the modest baseline performance; however, even a model with limited predictive power should exhibit a clear drop when its underlying causal logic is fundamentally reversed. Despite this consistent trend of insensitivity, we observed that Qwen2.5-512 exhibited slightly more sensitivity in certain conditions. Specifically, it showed a statistically significant decline under Reversed diff tags ($\Delta = 0.0126, p_{\text{Holm}} = 0.0272$) and a marginal drop under Spurious change markers ($\Delta = 0.0096, p_{\text{Holm}} = 0.3139$). In contrast, CodeBERT remained almost entirely unaffected by these same perturbations, suggesting that encoder-based architectures may be even more reliant on shallow, invariant cues than their decoder-based counterparts. However, even for the larger decoder-based model, the absolute magnitude of these drops remained remarkably small, further confirming that its predictive utility still relies heavily on surface-level regularities that persist under perturbation. These results highlight a critical semantic gap: current CLMs achieve stability not through robust modeling of code-change logic, but by falling back on shallow features. This underscores the urgent need for architectures or training objectives specifically designed to capture the relational dynamics of code transformations.

\begin{table*}[t]
\centering
\caption{Stage 2 perturbation analysis with CodeBERT and Qwen2.5-512 (mean $\pm$ std over 10 runs).}
{\footnotesize Percent change is relative to the corresponding original encoding. \par}
\label{tab:rq3_full}
\resizebox{\textwidth}{!}{%
\begin{tabular}{llccccc}
\toprule
Model & Condition & F1 & PR-AUC & Accuracy & Precision & Recall \\
\midrule
CodeBERT & After+Markers
         & $0.3524 \pm 0.0238$ & $0.3107 \pm 0.0202$ & $0.6152 \pm 0.0577$ & $0.2968 \pm 0.0175$ & $0.4489 \pm 0.1046$ \\
         & \emph{Spurious CHG markers}
         & $0.3532 \pm 0.0291$ & $0.3119 \pm 0.0328$ & $0.5830 \pm 0.1134$ & $0.2900 \pm 0.0225$ & $0.4943 \pm 0.1853$ \\
         & & {\scriptsize +0.23\%} \\
\midrule
CodeBERT & Before+After
         & $0.3521 \pm 0.0385$ & $0.3161 \pm 0.0101$ & $0.6161 \pm 0.0553$ & $0.2999 \pm 0.0197$ & $0.4536 \pm 0.1236$ \\
         & \emph{Swapped snapshots}
         & $0.3538 \pm 0.0381$ & $0.3178 \pm 0.0095$ & $0.6164 \pm 0.0556$ & $0.3012 \pm 0.0220$ & $0.4565 \pm 0.1236$ \\
         & & {\scriptsize +0.48\%} \\
\midrule
CodeBERT & Diff w/ tags
         & $0.3953 \pm 0.0100$ & $0.3349 \pm 0.0076$ & $0.5987 \pm 0.0346$ & $0.3082 \pm 0.0131$ & $0.5571 \pm 0.0602$ \\
         & \emph{Reversed diff tags}
         & $0.3919 \pm 0.0112$ & $0.3345 \pm 0.0079$ & $0.5999 \pm 0.0348$ & $0.3071 \pm 0.0131$ & $0.5476 \pm 0.0628$ \\
         & & {\scriptsize $-0.86$\%} \\

\midrule
CodeBERT & Added$\rightarrow$Deleted
         & $0.3971 \pm 0.0114$ & $0.3240 \pm 0.0106$ & $0.5988 \pm 0.0350$ & $0.3093 \pm 0.0156$ & $0.5609 \pm 0.0606$ \\
         & \emph{Swapped added/deleted blocks}
         & $0.3953 \pm 0.0095$ & $0.3299 \pm 0.0117$ & $0.5868 \pm 0.0432$ & $0.3042 \pm 0.0148$ & $0.5741 \pm 0.0749$ \\
         & & {\scriptsize $-0.45$\%} \\
\midrule
\midrule
Qwen2.5-512 & After+Markers
         & $0.3424 \pm 0.0109$ & --- & $0.4560 \pm 0.0363$ & $0.2400 \pm 0.0090$ & $0.6009 \pm 0.0485$ \\
         & \emph{Spurious CHG markers}
         & $0.3328 \pm 0.0218$ & --- & $0.4717 \pm 0.0261$ & $0.2372 \pm 0.0116$ & $0.5612 \pm 0.0704$ \\
         & & {\scriptsize $-2.80$\%} \\
\midrule
Qwen2.5-512  & Before+After
         & $0.3382 \pm 0.0130$ & --- & $0.4680 \pm 0.0403$ & $0.2401 \pm 0.0116$ & $0.5767 \pm 0.0528$ \\
         & \emph{Swapped snapshots}
         & $0.3392 \pm 0.0130$ & --- & $0.4719 \pm 0.0392$ & $0.2412 \pm 0.0099$ & $0.5754 \pm 0.0559$ \\
         & & {\scriptsize +0.30\%} \\
\midrule
Qwen2.5-512  & Diff w/ tags
         & $0.3536 \pm 0.0101$ & --- & $0.5115 \pm 0.0385$ & $0.2579 \pm 0.0114$ & $0.5666 \pm 0.0457$ \\
         & \emph{Reversed diff tags}
         & $0.3410 \pm 0.0125$ & --- & $0.4752 \pm 0.0300$ & $0.2426 \pm 0.0078$ & $0.5766 \pm 0.0503$ \\
         & & {\scriptsize $-3.56$\%} \\
\midrule
Qwen2.5-512   & Added$\rightarrow$Deleted
         & $0.3498 \pm 0.0139$ & --- & $0.4973 \pm 0.0468$ & $0.2528 \pm 0.0139$ & $0.5732 \pm 0.0574$ \\
         & \emph{Swapped blocks}
         & $0.3504 \pm 0.0117$ & --- & $0.4894 \pm 0.0407$ & $0.2511 \pm 0.0099$ & $0.5845 \pm 0.0571$ \\
         & & {\scriptsize +0.17\%} \\
\bottomrule
\end{tabular}
}
\end{table*}

\begin{tcolorbox}[colback=gray!10, colframe=gray!40, title=Summary of RQ3]
Both encoder and decoder-based models exhibited ``semantic blindness,'' with F1 remaining stable even when code-change logic or polarity was reversed ($\Delta \leq 1.26\%$). While Qwen2.5 showed marginal sensitivity to template tags, CodeBERT was almost entirely unaffected by any perturbations. This pervasive robustness suggests that current CLMs rely on shallow, invariant surface cues rather than a genuine understanding of code-change semantics.
\end{tcolorbox}

\section{Threats to Validity}

\subsection{Internal Validity}
\noindent\textbf{Subjectivity in Manual Verification.} Determining whether a code modification is actually bug-inducing is an inherently difficult and subjective task. To mitigate this subjectivity, our manual audit process relied primarily on explicit commit-message evidence (e.g., statements such as ``reverting due to a crash'') rather than a subjective interpretation of code logic. This is because such natural-language mentions provide the most direct and reliable justification from the developers who performed the revert. In cases where the commit messages were vague or lacked sufficient reasoning for the revert, we categorized them as incorrectly collected samples. Consequently, this conservative auditing approach revealed an error rate of only 8\% (Table~\ref{tab:rq1-audit}), thereby demonstrating the high precision of our curated dataset. Despite these rigorous criteria, human inspection is not infallible, and the potential for misjudgment during the verification of the 100 samples remains a threat to internal validity.
 
\noindent\textbf{Input Encodings and Perturbation Strategies.} We evaluated five representative encodings, including formats inspired by prior JIT-SDP studies. However, all encodings in our study are derived solely from \texttt{function\_before} and \texttt{function\_after}, which prevents us from leveraging richer semantic signals available in other sources such as commit messages. Moreover, even within these code-centric views, alternative ways of structuring the information---such as tree-structured or graph-based encodings---may provide richer signals of semantic change that are not fully captured by the linear, token-based representations used in this work. Additionally, we designed four perturbations as diagnostic probes specifically intended to assess semantic sensitivity. While our results demonstrated that models remain largely indifferent even under significant semantic disruptions, there may exist alternative perturbation strategies that could further expose the extent of this insensitivity or uncover different facets of modification reasoning. Exploring such alternatives and their alignment with diverse real-world scenarios remains an open avenue for future work.

\noindent\textbf{Statistical Power.} All statistical tests were based on 10 independent runs with different seeds. While this design controls stochastic variance, a larger number of repetitions could yield stronger statistical power and narrower confidence intervals.
However, prior AI studies on code intelligence tasks rarely report extensive repetitions: many omit random seeds entirely or rely on a single run, and even the more rigorous works seldom exceed 10 repetitions due to the substantial computational cost of CLMs~\cite{ye2023tram,grishina2023earlybird}. Running 30 or more trials per configuration is generally infeasible in practice. Our choice of 10 runs follows this common practice while still providing a reasonable trade-off between robustness and resource constraints.

\subsection{External Validity}
\noindent\textbf{Dataset Coverage and Generalizability.} We intentionally restricted the scope of ReDef to single-function modifications anchored by revert commits. Since reverts and single-function edits represent only a subset of all defective modifications, our dataset does not encompass the full spectrum of defect patterns, such as multi-function changes or bugs that are not easily identified for reversion. This focus may limit the generalizability of our findings to more complex, ``tangled'' commits. However, this conservative approach significantly mitigates the ``blame-innocent'' problem inherent in traditional SZZ-based labeling, ensuring high label precision. By prioritizing reliability over sheer volume, ReDef serves as a high-confidence benchmark that complements existing large-scale but noisy datasets, providing a necessary trade-off between coverage and predictive reliability.

\noindent\textbf{Project Selection Bias.} Our dataset was drawn from 22 large-scale open-source projects across diverse domains. However, the data volume is notably concentrated in a few massive repositories, particularly Linux and Chromium. This skew risks biasing our findings toward system-level software patterns, potentially limiting their generalizability to smaller applications or projects with different review and revert policies. Furthermore, industrial or proprietary projects could exhibit markedly different defect patterns, and including such settings in future work would provide stronger external validity.

\noindent\textbf{Cross-Language Applicability.} This study focused solely on C/C++ projects. Substantial syntactic and semantic differences across programming languages may limit the applicability of our findings. Extending evaluation to multi-language datasets is a key direction for future work. Such efforts should examine whether our observations hold consistently across different language families.

\section{Conclusion}

This work introduced ReDef, a high-confidence JIT-SDP dataset consisting of 3,164 defective and 10,268 clean function-level modifications. Manual audits confirmed a label precision of 92\%, which is substantially higher than that of prior JIT-SDP corpora. Our curation pipeline---combining revert anchoring, longitudinal history screening, and GPT-assisted triage---effectively minimizes the label noise prevalent in existing datasets. By overcoming the limitations of conventional SZZ-based resources, ReDef provides a robust foundation for the research community to evaluate change-aware models. This benchmark not only enables reproducible comparisons but also facilitates the development of next-generation, edit-sensitive learning objectives.

Using ReDef, we evaluated four representative Code Language Models (CLMs)---CodeBERT, CodeT5+, UniXcoder, and Qwen2.5---across five input encoding strategies. Our results indicate that compact, change-focused encodings consistently achieve the best performance by emphasizing localized modifications and minimizing information loss due to truncation. Notably, we observed that larger context windows do not necessarily translate to better performance; both UniXcoder and Qwen2.5 performed worse under a 1,024-token budget than at 512 tokens, suggesting that excessive context can obscure critical change signals.

Counterfactual perturbations of change representations further revealed that CLMs' performance remains effectively stable even under significant structural distortions. While encoder-based models were almost entirely unaffected by these perturbations, the decoder-based Qwen2.5 exhibited only marginal sensitivity in specific conditions. Even when we reversed the logical order of modifications or injected spurious change markers, F1 remained nearly identical. If these models had truly internalized the semantics of code edits, such fundamental distortions should have led to a precipitous decline in predictive performance.

This outcome exposes a critical blind spot in current evaluations of code-change models: what appears to be robustness is, in fact, a symptom of semantic blindness, where models fall back on surface-level cues rather than genuine understanding. Our findings caution against interpreting high benchmark scores as evidence of an advanced grasp of code evolution. Ultimately, this study underscores the need for new architectures and training objectives specifically designed to capture the relational and causal dynamics of code transformations.

Several threats to validity remain. While ReDef ensures high label precision, its focus on single-function reverts in large-scale C/C++ projects may limit generalizability to multi-function commits or other software domains. Furthermore, despite our manual audits, the inherent subjectivity of human inspection persists as a potential bias, and alternative encoding or perturbation strategies may exist that could uncover deeper semantic insights. Future research should explore the underlying causes of this semantic lag, potentially through edit-aware pre-training objectives or architectures specifically tailored for change reasoning. It is equally vital to investigate whether this reliance on superficial patterns persists in other semantics-sensitive tasks, such as vulnerability detection and automated code review. Finally, evaluating larger API-based models (e.g., GPT-4o) under the same protocol will reveal whether increased scale can effectively bridge the critical ``semantic gap'' exposed in this study.

We hope that ReDef and our evaluation protocol will serve as a foundation for the next generation of change-aware, semantics-sensitive defect prediction models.

\section*{Data Availability}
An anonymous link containing the dataset, construction scripts, and experimental code is publicly available for reproducibility:\url{https://figshare.com/s/4f202bc0921e26b41dc2}.

%
%
%

\bibliographystyle{plainnat}
\bibliography{references}


\begin{thebibliography}{68}


\ifx \showCODEN    \undefined \def \showCODEN     #1{\unskip}     \fi
\ifx \showISBNx    \undefined \def \showISBNx     #1{\unskip}     \fi
\ifx \showISBNxiii \undefined \def \showISBNxiii  #1{\unskip}     \fi
\ifx \showISSN     \undefined \def \showISSN      #1{\unskip}     \fi
\ifx \showLCCN     \undefined \def \showLCCN      #1{\unskip}     \fi
\ifx \shownote     \undefined \def \shownote      #1{#1}          \fi
\ifx \showarticletitle \undefined \def \showarticletitle #1{#1}   \fi
\ifx \showURL      \undefined \def \showURL       {\relax}        \fi
\providecommand\bibfield[2]{#2}
\providecommand\bibinfo[2]{#2}
\providecommand\natexlab[1]{#1}
\providecommand\showeprint[2][]{arXiv:#2}

\bibitem[Abu~Talib et~al\mbox{.}(2024)]%
        {abu2024parameter}
\bibfield{author}{\bibinfo{person}{Manar Abu~Talib}, \bibinfo{person}{Ali Bou~Nassif}, \bibinfo{person}{Mohammad Azzeh}, \bibinfo{person}{Yaser Alesh}, {and} \bibinfo{person}{Yaman Afadar}.} \bibinfo{year}{2024}\natexlab{}.
\newblock \showarticletitle{Parameter-efficient fine-tuning of pre-trained code models for just-in-time defect prediction}.
\newblock \bibinfo{journal}{\emph{Neural Computing and Applications}} \bibinfo{volume}{36}, \bibinfo{number}{27} (\bibinfo{year}{2024}), \bibinfo{pages}{16911--16940}.
\newblock


\bibitem[Afric et~al\mbox{.}(2023)]%
        {afric2023empirical}
\bibfield{author}{\bibinfo{person}{Petar Afric}, \bibinfo{person}{Davor Vukadin}, \bibinfo{person}{Marin Silic}, {and} \bibinfo{person}{Goran Delac}.} \bibinfo{year}{2023}\natexlab{}.
\newblock \showarticletitle{Empirical study: How issue classification influences software defect prediction}.
\newblock \bibinfo{journal}{\emph{IEEE access}}  \bibinfo{volume}{11} (\bibinfo{year}{2023}), \bibinfo{pages}{11732--11748}.
\newblock


\bibitem[Balaram and Vasundra(2022)]%
        {balaram2022prediction}
\bibfield{author}{\bibinfo{person}{A Balaram} {and} \bibinfo{person}{S Vasundra}.} \bibinfo{year}{2022}\natexlab{}.
\newblock \showarticletitle{Prediction of software fault-prone classes using ensemble random forest with adaptive synthetic sampling algorithm}.
\newblock \bibinfo{journal}{\emph{Automated Software Engineering}} \bibinfo{volume}{29}, \bibinfo{number}{1} (\bibinfo{year}{2022}), \bibinfo{pages}{6}.
\newblock


\bibitem[Bell et~al\mbox{.}(2013)]%
        {bell2013limited}
\bibfield{author}{\bibinfo{person}{Robert~M Bell}, \bibinfo{person}{Thomas~J Ostrand}, {and} \bibinfo{person}{Elaine~J Weyuker}.} \bibinfo{year}{2013}\natexlab{}.
\newblock \showarticletitle{The limited impact of individual developer data on software defect prediction}.
\newblock \bibinfo{journal}{\emph{Empirical Software Engineering}} \bibinfo{volume}{18}, \bibinfo{number}{3} (\bibinfo{year}{2013}), \bibinfo{pages}{478--505}.
\newblock


\bibitem[Buda et~al\mbox{.}(2018)]%
        {buda2018systematic}
\bibfield{author}{\bibinfo{person}{Mateusz Buda}, \bibinfo{person}{Atsuto Maki}, {and} \bibinfo{person}{Maciej~A Mazurowski}.} \bibinfo{year}{2018}\natexlab{}.
\newblock \showarticletitle{A systematic study of the class imbalance problem in convolutional neural networks}.
\newblock \bibinfo{journal}{\emph{Neural networks}}  \bibinfo{volume}{106} (\bibinfo{year}{2018}), \bibinfo{pages}{249--259}.
\newblock


\bibitem[Chen et~al\mbox{.}(2023)]%
        {chen2023diversevul}
\bibfield{author}{\bibinfo{person}{Yizheng Chen}, \bibinfo{person}{Zhoujie Ding}, \bibinfo{person}{Lamya Alowain}, \bibinfo{person}{Xinyun Chen}, {and} \bibinfo{person}{David Wagner}.} \bibinfo{year}{2023}\natexlab{}.
\newblock \showarticletitle{Diversevul: A new vulnerable source code dataset for deep learning based vulnerability detection}. In \bibinfo{booktitle}{\emph{Proceedings of the 26th International Symposium on Research in Attacks, Intrusions and Defenses}}. \bibinfo{pages}{654--668}.
\newblock


\bibitem[Cohen(2013)]%
        {cohen2013statistical}
\bibfield{author}{\bibinfo{person}{Jacob Cohen}.} \bibinfo{year}{2013}\natexlab{}.
\newblock \bibinfo{booktitle}{\emph{Statistical power analysis for the behavioral sciences}}.
\newblock \bibinfo{publisher}{routledge}.
\newblock


\bibitem[Da~Costa et~al\mbox{.}(2016)]%
        {da2016framework}
\bibfield{author}{\bibinfo{person}{Daniel~Alencar Da~Costa}, \bibinfo{person}{Shane McIntosh}, \bibinfo{person}{Weiyi Shang}, \bibinfo{person}{Uir{\'a} Kulesza}, \bibinfo{person}{Roberta Coelho}, {and} \bibinfo{person}{Ahmed~E Hassan}.} \bibinfo{year}{2016}\natexlab{}.
\newblock \showarticletitle{A framework for evaluating the results of the szz approach for identifying bug-introducing changes}.
\newblock \bibinfo{journal}{\emph{IEEE Transactions on Software Engineering}} \bibinfo{volume}{43}, \bibinfo{number}{7} (\bibinfo{year}{2016}), \bibinfo{pages}{641--657}.
\newblock


\bibitem[Dahou et~al\mbox{.}(2025)]%
        {dahou2025semantic}
\bibfield{author}{\bibinfo{person}{Abdelhalim Dahou}, \bibinfo{person}{Ansgar Scherp}, \bibinfo{person}{Sebastian Kurten}, \bibinfo{person}{Brigitte Mathiak}, {and} \bibinfo{person}{Madhu Chauhan}.} \bibinfo{year}{2025}\natexlab{}.
\newblock \showarticletitle{Semantic Source Code Segmentation using Small and Large Language Models}.
\newblock \bibinfo{journal}{\emph{arXiv preprint arXiv:2507.08992}} (\bibinfo{year}{2025}).
\newblock


\bibitem[Daniel~Han and team(2023)]%
        {unsloth}
\bibfield{author}{\bibinfo{person}{Michael~Han Daniel~Han} {and} \bibinfo{person}{Unsloth team}.} \bibinfo{year}{2023}\natexlab{}.
\newblock \bibinfo{booktitle}{\emph{Unsloth}}.
\newblock
\urldef\tempurl%
\url{http://github.com/unslothai/unsloth}
\showURL{%
\tempurl}


\bibitem[Ding et~al\mbox{.}(2024)]%
        {ding2024vulnerability}
\bibfield{author}{\bibinfo{person}{Yangruibo Ding}, \bibinfo{person}{Yanjun Fu}, \bibinfo{person}{Omniyyah Ibrahim}, \bibinfo{person}{Chawin Sitawarin}, \bibinfo{person}{Xinyun Chen}, \bibinfo{person}{Basel Alomair}, \bibinfo{person}{David Wagner}, \bibinfo{person}{Baishakhi Ray}, {and} \bibinfo{person}{Yizheng Chen}.} \bibinfo{year}{2024}\natexlab{}.
\newblock \showarticletitle{Vulnerability detection with code language models: How far are we?}
\newblock \bibinfo{journal}{\emph{arXiv preprint arXiv:2403.18624}} (\bibinfo{year}{2024}).
\newblock


\bibitem[Elish and Elish(2008)]%
        {elish2008predicting}
\bibfield{author}{\bibinfo{person}{Karim~O Elish} {and} \bibinfo{person}{Mahmoud~O Elish}.} \bibinfo{year}{2008}\natexlab{}.
\newblock \showarticletitle{Predicting defect-prone software modules using support vector machines}.
\newblock \bibinfo{journal}{\emph{Journal of Systems and Software}} \bibinfo{volume}{81}, \bibinfo{number}{5} (\bibinfo{year}{2008}), \bibinfo{pages}{649--660}.
\newblock


\bibitem[Fan et~al\mbox{.}(2020)]%
        {fan2020ac}
\bibfield{author}{\bibinfo{person}{Jiahao Fan}, \bibinfo{person}{Yi Li}, \bibinfo{person}{Shaohua Wang}, {and} \bibinfo{person}{Tien~N Nguyen}.} \bibinfo{year}{2020}\natexlab{}.
\newblock \showarticletitle{AC/C++ code vulnerability dataset with code changes and CVE summaries}. In \bibinfo{booktitle}{\emph{Proceedings of the 17th international conference on mining software repositories}}. \bibinfo{pages}{508--512}.
\newblock


\bibitem[Fan et~al\mbox{.}(2019)]%
        {fan2019impact}
\bibfield{author}{\bibinfo{person}{Yuanrui Fan}, \bibinfo{person}{Xin Xia}, \bibinfo{person}{Daniel~Alencar Da~Costa}, \bibinfo{person}{David Lo}, \bibinfo{person}{Ahmed~E Hassan}, {and} \bibinfo{person}{Shanping Li}.} \bibinfo{year}{2019}\natexlab{}.
\newblock \showarticletitle{The impact of mislabeled changes by szz on just-in-time defect prediction}.
\newblock \bibinfo{journal}{\emph{IEEE transactions on software engineering}} \bibinfo{volume}{47}, \bibinfo{number}{8} (\bibinfo{year}{2019}), \bibinfo{pages}{1559--1586}.
\newblock


\bibitem[Feng et~al\mbox{.}(2020)]%
        {feng2020codebert}
\bibfield{author}{\bibinfo{person}{Zhangyin Feng}, \bibinfo{person}{Daya Guo}, \bibinfo{person}{Duyu Tang}, \bibinfo{person}{Nan Duan}, \bibinfo{person}{Xiaocheng Feng}, \bibinfo{person}{Ming Gong}, \bibinfo{person}{Linjun Shou}, \bibinfo{person}{Bing Qin}, \bibinfo{person}{Ting Liu}, \bibinfo{person}{Daxin Jiang}, {et~al\mbox{.}}} \bibinfo{year}{2020}\natexlab{}.
\newblock \showarticletitle{Codebert: A pre-trained model for programming and natural languages}.
\newblock \bibinfo{journal}{\emph{arXiv preprint arXiv:2002.08155}} (\bibinfo{year}{2020}).
\newblock


\bibitem[Gao et~al\mbox{.}(2024)]%
        {gao2024learning}
\bibfield{author}{\bibinfo{person}{Shuzheng Gao}, \bibinfo{person}{Wenxin Mao}, \bibinfo{person}{Cuiyun Gao}, \bibinfo{person}{Li Li}, \bibinfo{person}{Xing Hu}, \bibinfo{person}{Xin Xia}, {and} \bibinfo{person}{Michael~R Lyu}.} \bibinfo{year}{2024}\natexlab{}.
\newblock \showarticletitle{Learning in the wild: Towards leveraging unlabeled data for effectively tuning pre-trained code models}. In \bibinfo{booktitle}{\emph{Proceedings of the IEEE/ACM 46th International Conference on Software Engineering}}. \bibinfo{pages}{1--13}.
\newblock


\bibitem[Grattafiori et~al\mbox{.}(2024)]%
        {grattafiori2024llama}
\bibfield{author}{\bibinfo{person}{Aaron Grattafiori}, \bibinfo{person}{Abhimanyu Dubey}, \bibinfo{person}{Abhinav Jauhri}, \bibinfo{person}{Abhinav Pandey}, \bibinfo{person}{Abhishek Kadian}, \bibinfo{person}{Ahmad Al-Dahle}, \bibinfo{person}{Aiesha Letman}, \bibinfo{person}{Akhil Mathur}, \bibinfo{person}{Alan Schelten}, \bibinfo{person}{Alex Vaughan}, {et~al\mbox{.}}} \bibinfo{year}{2024}\natexlab{}.
\newblock \showarticletitle{The llama 3 herd of models}.
\newblock \bibinfo{journal}{\emph{arXiv preprint arXiv:2407.21783}} (\bibinfo{year}{2024}).
\newblock


\bibitem[Grishina et~al\mbox{.}(2023)]%
        {grishina2023earlybird}
\bibfield{author}{\bibinfo{person}{Anastasiia Grishina}, \bibinfo{person}{Max Hort}, {and} \bibinfo{person}{Leon Moonen}.} \bibinfo{year}{2023}\natexlab{}.
\newblock \showarticletitle{The earlybird catches the bug: On exploiting early layers of encoder models for more efficient code classification}. In \bibinfo{booktitle}{\emph{Proceedings of the 31st ACM Joint European Software Engineering Conference and Symposium on the Foundations of Software Engineering}}. \bibinfo{pages}{895--907}.
\newblock


\bibitem[Guo et~al\mbox{.}(2022)]%
        {guo2022unixcoder}
\bibfield{author}{\bibinfo{person}{Daya Guo}, \bibinfo{person}{Shuai Lu}, \bibinfo{person}{Nan Duan}, \bibinfo{person}{Yanlin Wang}, \bibinfo{person}{Ming Zhou}, {and} \bibinfo{person}{Jian Yin}.} \bibinfo{year}{2022}\natexlab{}.
\newblock \showarticletitle{Unixcoder: Unified cross-modal pre-training for code representation}.
\newblock \bibinfo{journal}{\emph{arXiv preprint arXiv:2203.03850}} (\bibinfo{year}{2022}).
\newblock


\bibitem[Guo et~al\mbox{.}(2023)]%
        {guo2023study}
\bibfield{author}{\bibinfo{person}{Yuxiang Guo}, \bibinfo{person}{Xiaopeng Gao}, \bibinfo{person}{Zhenyu Zhang}, \bibinfo{person}{Wing~Kwong Chan}, {and} \bibinfo{person}{Bo Jiang}.} \bibinfo{year}{2023}\natexlab{}.
\newblock \showarticletitle{A study on the impact of pre-trained model on Just-In-Time defect prediction}. In \bibinfo{booktitle}{\emph{2023 IEEE 23rd International Conference on Software Quality, Reliability, and Security (QRS)}}. IEEE, \bibinfo{pages}{105--116}.
\newblock


\bibitem[Herbold et~al\mbox{.}(2022)]%
        {herbold2022problems}
\bibfield{author}{\bibinfo{person}{Steffen Herbold}, \bibinfo{person}{Alexander Trautsch}, \bibinfo{person}{Fabian Trautsch}, {and} \bibinfo{person}{Benjamin Ledel}.} \bibinfo{year}{2022}\natexlab{}.
\newblock \showarticletitle{Problems with SZZ and features: An empirical study of the state of practice of defect prediction data collection}.
\newblock \bibinfo{journal}{\emph{Empirical Software Engineering}} \bibinfo{volume}{27}, \bibinfo{number}{2} (\bibinfo{year}{2022}), \bibinfo{pages}{42}.
\newblock


\bibitem[Hoang et~al\mbox{.}(2019)]%
        {hoang2019deepjit}
\bibfield{author}{\bibinfo{person}{Thong Hoang}, \bibinfo{person}{Hoa~Khanh Dam}, \bibinfo{person}{Yasutaka Kamei}, \bibinfo{person}{David Lo}, {and} \bibinfo{person}{Naoyasu Ubayashi}.} \bibinfo{year}{2019}\natexlab{}.
\newblock \showarticletitle{Deepjit: an end-to-end deep learning framework for just-in-time defect prediction}. In \bibinfo{booktitle}{\emph{2019 IEEE/ACM 16th International Conference on Mining Software Repositories (MSR)}}. IEEE, \bibinfo{pages}{34--45}.
\newblock


\bibitem[Hoang et~al\mbox{.}(2020)]%
        {hoang2020cc2vec}
\bibfield{author}{\bibinfo{person}{Thong Hoang}, \bibinfo{person}{Hong~Jin Kang}, \bibinfo{person}{David Lo}, {and} \bibinfo{person}{Julia Lawall}.} \bibinfo{year}{2020}\natexlab{}.
\newblock \showarticletitle{Cc2vec: Distributed representations of code changes}. In \bibinfo{booktitle}{\emph{Proceedings of the ACM/IEEE 42nd international conference on software engineering}}. \bibinfo{pages}{518--529}.
\newblock


\bibitem[Holm(1979)]%
        {holm1979simple}
\bibfield{author}{\bibinfo{person}{Sture Holm}.} \bibinfo{year}{1979}\natexlab{}.
\newblock \showarticletitle{A simple sequentially rejective multiple test procedure}.
\newblock \bibinfo{journal}{\emph{Scandinavian journal of statistics}} (\bibinfo{year}{1979}), \bibinfo{pages}{65--70}.
\newblock


\bibitem[Huda et~al\mbox{.}(2017)]%
        {huda2017framework}
\bibfield{author}{\bibinfo{person}{Shamsul Huda}, \bibinfo{person}{Sultan Alyahya}, \bibinfo{person}{Md~Mohsin Ali}, \bibinfo{person}{Shafiq Ahmad}, \bibinfo{person}{Jemal Abawajy}, \bibinfo{person}{Hmood Al-Dossari}, {and} \bibinfo{person}{John Yearwood}.} \bibinfo{year}{2017}\natexlab{}.
\newblock \showarticletitle{A framework for software defect prediction and metric selection}.
\newblock \bibinfo{journal}{\emph{IEEE access}}  \bibinfo{volume}{6} (\bibinfo{year}{2017}), \bibinfo{pages}{2844--2858}.
\newblock


\bibitem[Hui et~al\mbox{.}(2024)]%
        {hui2024qwen2}
\bibfield{author}{\bibinfo{person}{Binyuan Hui}, \bibinfo{person}{Jian Yang}, \bibinfo{person}{Zeyu Cui}, \bibinfo{person}{Jiaxi Yang}, \bibinfo{person}{Dayiheng Liu}, \bibinfo{person}{Lei Zhang}, \bibinfo{person}{Tianyu Liu}, \bibinfo{person}{Jiajun Zhang}, \bibinfo{person}{Bowen Yu}, \bibinfo{person}{Keming Lu}, {et~al\mbox{.}}} \bibinfo{year}{2024}\natexlab{}.
\newblock \showarticletitle{Qwen2. 5-coder technical report}.
\newblock \bibinfo{journal}{\emph{arXiv preprint arXiv:2409.12186}} (\bibinfo{year}{2024}).
\newblock


\bibitem[Jia and Liang(2017)]%
        {jia2017adversarial}
\bibfield{author}{\bibinfo{person}{Robin Jia} {and} \bibinfo{person}{Percy Liang}.} \bibinfo{year}{2017}\natexlab{}.
\newblock \showarticletitle{Adversarial examples for evaluating reading comprehension systems}.
\newblock \bibinfo{journal}{\emph{arXiv preprint arXiv:1707.07328}} (\bibinfo{year}{2017}).
\newblock


\bibitem[Jiang et~al\mbox{.}(2025)]%
        {jiang2025just}
\bibfield{author}{\bibinfo{person}{Yuze Jiang}, \bibinfo{person}{Beijun Shen}, {and} \bibinfo{person}{Xiaodong Gu}.} \bibinfo{year}{2025}\natexlab{}.
\newblock \showarticletitle{Just-in-time software defect prediction via bi-modal change representation learning}.
\newblock \bibinfo{journal}{\emph{Journal of Systems and Software}}  \bibinfo{volume}{219} (\bibinfo{year}{2025}), \bibinfo{pages}{112253}.
\newblock


\bibitem[Just et~al\mbox{.}(2014)]%
        {just2014defects4j}
\bibfield{author}{\bibinfo{person}{Ren{\'e} Just}, \bibinfo{person}{Darioush Jalali}, {and} \bibinfo{person}{Michael~D Ernst}.} \bibinfo{year}{2014}\natexlab{}.
\newblock \showarticletitle{Defects4J: A database of existing faults to enable controlled testing studies for Java programs}. In \bibinfo{booktitle}{\emph{Proceedings of the 2014 international symposium on software testing and analysis}}. \bibinfo{pages}{437--440}.
\newblock


\bibitem[Kamei et~al\mbox{.}(2016)]%
        {kamei2016studying}
\bibfield{author}{\bibinfo{person}{Yasutaka Kamei}, \bibinfo{person}{Takafumi Fukushima}, \bibinfo{person}{Shane McIntosh}, \bibinfo{person}{Kazuhiro Yamashita}, \bibinfo{person}{Naoyasu Ubayashi}, {and} \bibinfo{person}{Ahmed~E Hassan}.} \bibinfo{year}{2016}\natexlab{}.
\newblock \showarticletitle{Studying just-in-time defect prediction using cross-project models}.
\newblock \bibinfo{journal}{\emph{Empirical Software Engineering}} \bibinfo{volume}{21}, \bibinfo{number}{5} (\bibinfo{year}{2016}), \bibinfo{pages}{2072--2106}.
\newblock


\bibitem[Kamei et~al\mbox{.}(2012)]%
        {kamei2012large}
\bibfield{author}{\bibinfo{person}{Yasutaka Kamei}, \bibinfo{person}{Emad Shihab}, \bibinfo{person}{Bram Adams}, \bibinfo{person}{Ahmed~E Hassan}, \bibinfo{person}{Audris Mockus}, \bibinfo{person}{Anand Sinha}, {and} \bibinfo{person}{Naoyasu Ubayashi}.} \bibinfo{year}{2012}\natexlab{}.
\newblock \showarticletitle{A large-scale empirical study of just-in-time quality assurance}.
\newblock \bibinfo{journal}{\emph{IEEE Transactions on Software Engineering}} \bibinfo{volume}{39}, \bibinfo{number}{6} (\bibinfo{year}{2012}), \bibinfo{pages}{757--773}.
\newblock


\bibitem[Khoshgoftaar and Seliya(2002)]%
        {khoshgoftaar2002tree}
\bibfield{author}{\bibinfo{person}{Taghi~M Khoshgoftaar} {and} \bibinfo{person}{Naeem Seliya}.} \bibinfo{year}{2002}\natexlab{}.
\newblock \showarticletitle{Tree-based software quality estimation models for fault prediction}. In \bibinfo{booktitle}{\emph{Proceedings Eighth IEEE Symposium on Software Metrics}}. IEEE, \bibinfo{pages}{203--214}.
\newblock


\bibitem[Kim et~al\mbox{.}(2008)]%
        {kim2008classifying}
\bibfield{author}{\bibinfo{person}{Sunghun Kim}, \bibinfo{person}{E~James Whitehead}, {and} \bibinfo{person}{Yi Zhang}.} \bibinfo{year}{2008}\natexlab{}.
\newblock \showarticletitle{Classifying software changes: Clean or buggy?}
\newblock \bibinfo{journal}{\emph{IEEE Transactions on software engineering}} \bibinfo{volume}{34}, \bibinfo{number}{2} (\bibinfo{year}{2008}), \bibinfo{pages}{181--196}.
\newblock


\bibitem[Kim et~al\mbox{.}(2006)]%
        {kim2006automatic}
\bibfield{author}{\bibinfo{person}{Sunghun Kim}, \bibinfo{person}{Thomas Zimmermann}, \bibinfo{person}{Kai Pan}, \bibinfo{person}{E James~Jr}, {et~al\mbox{.}}} \bibinfo{year}{2006}\natexlab{}.
\newblock \showarticletitle{Automatic identification of bug-introducing changes}. In \bibinfo{booktitle}{\emph{21st IEEE/ACM international conference on automated software engineering (ASE'06)}}. IEEE, \bibinfo{pages}{81--90}.
\newblock


\bibitem[King and Zeng(2001)]%
        {king2001logistic}
\bibfield{author}{\bibinfo{person}{Gary King} {and} \bibinfo{person}{Langche Zeng}.} \bibinfo{year}{2001}\natexlab{}.
\newblock \showarticletitle{Logistic regression in rare events data}.
\newblock \bibinfo{journal}{\emph{Political analysis}} \bibinfo{volume}{9}, \bibinfo{number}{2} (\bibinfo{year}{2001}), \bibinfo{pages}{137--163}.
\newblock


\bibitem[Lakens(2013)]%
        {lakens2013calculating}
\bibfield{author}{\bibinfo{person}{Dani{\"e}l Lakens}.} \bibinfo{year}{2013}\natexlab{}.
\newblock \showarticletitle{Calculating and reporting effect sizes to facilitate cumulative science: a practical primer for t-tests and ANOVAs}.
\newblock \bibinfo{journal}{\emph{Frontiers in psychology}}  \bibinfo{volume}{4} (\bibinfo{year}{2013}), \bibinfo{pages}{863}.
\newblock


\bibitem[Lin et~al\mbox{.}(2023)]%
        {lin2023cct5}
\bibfield{author}{\bibinfo{person}{Bo Lin}, \bibinfo{person}{Shangwen Wang}, \bibinfo{person}{Zhongxin Liu}, \bibinfo{person}{Yepang Liu}, \bibinfo{person}{Xin Xia}, {and} \bibinfo{person}{Xiaoguang Mao}.} \bibinfo{year}{2023}\natexlab{}.
\newblock \showarticletitle{Cct5: A code-change-oriented pre-trained model}. In \bibinfo{booktitle}{\emph{Proceedings of the 31st ACM Joint European Software Engineering Conference and Symposium on the Foundations of Software Engineering}}. \bibinfo{pages}{1509--1521}.
\newblock


\bibitem[Loshchilov and Hutter(2017)]%
        {loshchilov2017decoupled}
\bibfield{author}{\bibinfo{person}{Ilya Loshchilov} {and} \bibinfo{person}{Frank Hutter}.} \bibinfo{year}{2017}\natexlab{}.
\newblock \showarticletitle{Decoupled weight decay regularization}.
\newblock \bibinfo{journal}{\emph{arXiv preprint arXiv:1711.05101}} (\bibinfo{year}{2017}).
\newblock


\bibitem[Lyu et~al\mbox{.}(2024)]%
        {lyu2024evaluating}
\bibfield{author}{\bibinfo{person}{Yunbo Lyu}, \bibinfo{person}{Hong~Jin Kang}, \bibinfo{person}{Ratnadira Widyasari}, \bibinfo{person}{Julia Lawall}, {and} \bibinfo{person}{David Lo}.} \bibinfo{year}{2024}\natexlab{}.
\newblock \showarticletitle{Evaluating SZZ implementations: An empirical study on the linux kernel}.
\newblock \bibinfo{journal}{\emph{IEEE Transactions on Software Engineering}} \bibinfo{volume}{50}, \bibinfo{number}{9} (\bibinfo{year}{2024}), \bibinfo{pages}{2219--2239}.
\newblock


\bibitem[Malhotra(2015)]%
        {malhotra2015systematic}
\bibfield{author}{\bibinfo{person}{Ruchika Malhotra}.} \bibinfo{year}{2015}\natexlab{}.
\newblock \showarticletitle{A systematic review of machine learning techniques for software fault prediction}.
\newblock \bibinfo{journal}{\emph{Applied Soft Computing}}  \bibinfo{volume}{27} (\bibinfo{year}{2015}), \bibinfo{pages}{504--518}.
\newblock


\bibitem[Neto et~al\mbox{.}(2018)]%
        {neto2018impact}
\bibfield{author}{\bibinfo{person}{Edmilson~Campos Neto}, \bibinfo{person}{Daniel~Alencar Da~Costa}, {and} \bibinfo{person}{Uir{\'a} Kulesza}.} \bibinfo{year}{2018}\natexlab{}.
\newblock \showarticletitle{The impact of refactoring changes on the SZZ algorithm: An empirical study}. In \bibinfo{booktitle}{\emph{2018 IEEE 25th international conference on software analysis, evolution and reengineering (SANER)}}. IEEE, \bibinfo{pages}{380--390}.
\newblock


\bibitem[Ni et~al\mbox{.}(2023)]%
        {ni2023function}
\bibfield{author}{\bibinfo{person}{Chao Ni}, \bibinfo{person}{Xinrong Guo}, \bibinfo{person}{Yan Zhu}, \bibinfo{person}{Xiaodan Xu}, {and} \bibinfo{person}{Xiaohu Yang}.} \bibinfo{year}{2023}\natexlab{}.
\newblock \showarticletitle{Function-level vulnerability detection through fusing multi-modal knowledge}. In \bibinfo{booktitle}{\emph{2023 38th IEEE/ACM International Conference on Automated Software Engineering (ASE)}}. IEEE, \bibinfo{pages}{1911--1918}.
\newblock


\bibitem[Ni et~al\mbox{.}(2022)]%
        {ni2022best}
\bibfield{author}{\bibinfo{person}{Chao Ni}, \bibinfo{person}{Wei Wang}, \bibinfo{person}{Kaiwen Yang}, \bibinfo{person}{Xin Xia}, \bibinfo{person}{Kui Liu}, {and} \bibinfo{person}{David Lo}.} \bibinfo{year}{2022}\natexlab{}.
\newblock \showarticletitle{The best of both worlds: integrating semantic features with expert features for defect prediction and localization}. In \bibinfo{booktitle}{\emph{Proceedings of the 30th ACM Joint European Software Engineering Conference and Symposium on the Foundations of Software Engineering}}. \bibinfo{pages}{672--683}.
\newblock


\bibitem[Niu et~al\mbox{.}(2023)]%
        {niu2023empirical}
\bibfield{author}{\bibinfo{person}{Changan Niu}, \bibinfo{person}{Chuanyi Li}, \bibinfo{person}{Vincent Ng}, \bibinfo{person}{Dongxiao Chen}, \bibinfo{person}{Jidong Ge}, {and} \bibinfo{person}{Bin Luo}.} \bibinfo{year}{2023}\natexlab{}.
\newblock \showarticletitle{An empirical comparison of pre-trained models of source code}. In \bibinfo{booktitle}{\emph{2023 IEEE/ACM 45th International Conference on Software Engineering (ICSE)}}. IEEE, \bibinfo{pages}{2136--2148}.
\newblock


\bibitem[Niu et~al\mbox{.}(2025)]%
        {niu2025refactoring}
\bibfield{author}{\bibinfo{person}{Feifei Niu}, \bibinfo{person}{Junqian Shao}, \bibinfo{person}{Christoph Mayr-Dorn}, \bibinfo{person}{Liguo Huang}, \bibinfo{person}{Wesley~KG Assun{\c{c}}{\~a}o}, \bibinfo{person}{Chuanyi Li}, \bibinfo{person}{Jidong Ge}, {and} \bibinfo{person}{Alexander Egyed}.} \bibinfo{year}{2025}\natexlab{}.
\newblock \showarticletitle{Refactoring $\neq$ Bug-Inducing: Improving Defect Prediction with Code Change Tactics Analysis}.
\newblock \bibinfo{journal}{\emph{arXiv preprint arXiv:2507.19714}} (\bibinfo{year}{2025}).
\newblock


\bibitem[Omri and Sinz(2020)]%
        {omri2020deep}
\bibfield{author}{\bibinfo{person}{Safa Omri} {and} \bibinfo{person}{Carsten Sinz}.} \bibinfo{year}{2020}\natexlab{}.
\newblock \showarticletitle{Deep learning for software defect prediction: A survey}. In \bibinfo{booktitle}{\emph{Proceedings of the IEEE/ACM 42nd international conference on software engineering workshops}}. \bibinfo{pages}{209--214}.
\newblock


\bibitem[Ostrand and Weyuker(2007)]%
        {ostrand2007measure}
\bibfield{author}{\bibinfo{person}{Thomas~J Ostrand} {and} \bibinfo{person}{Elaine~J Weyuker}.} \bibinfo{year}{2007}\natexlab{}.
\newblock \showarticletitle{How to measure success of fault prediction models}. In \bibinfo{booktitle}{\emph{Fourth international workshop on Software quality assurance: in conjunction with the 6th ESEC/FSE joint meeting}}. \bibinfo{pages}{25--30}.
\newblock


\bibitem[Ribeiro et~al\mbox{.}(2018)]%
        {ribeiro2018semantically}
\bibfield{author}{\bibinfo{person}{Marco~Tulio Ribeiro}, \bibinfo{person}{Sameer Singh}, {and} \bibinfo{person}{Carlos Guestrin}.} \bibinfo{year}{2018}\natexlab{}.
\newblock \showarticletitle{Semantically equivalent adversarial rules for debugging NLP models}. In \bibinfo{booktitle}{\emph{Proceedings of the 56th Annual Meeting of the Association for Computational Linguistics (volume 1: long papers)}}. \bibinfo{pages}{856--865}.
\newblock


\bibitem[Shihab et~al\mbox{.}(2012)]%
        {shihab2012industrial}
\bibfield{author}{\bibinfo{person}{Emad Shihab}, \bibinfo{person}{Ahmed~E Hassan}, \bibinfo{person}{Bram Adams}, {and} \bibinfo{person}{Zhen~Ming Jiang}.} \bibinfo{year}{2012}\natexlab{}.
\newblock \showarticletitle{An industrial study on the risk of software changes}. In \bibinfo{booktitle}{\emph{Proceedings of the ACM SIGSOFT 20th International Symposium on the Foundations of Software Engineering}}. \bibinfo{pages}{1--11}.
\newblock


\bibitem[{\'S}liwerski et~al\mbox{.}(2005)]%
        {sliwerski2005changes}
\bibfield{author}{\bibinfo{person}{Jacek {\'S}liwerski}, \bibinfo{person}{Thomas Zimmermann}, {and} \bibinfo{person}{Andreas Zeller}.} \bibinfo{year}{2005}\natexlab{}.
\newblock \showarticletitle{When do changes induce fixes?}
\newblock \bibinfo{journal}{\emph{ACM sigsoft software engineering notes}} \bibinfo{volume}{30}, \bibinfo{number}{4} (\bibinfo{year}{2005}), \bibinfo{pages}{1--5}.
\newblock


\bibitem[Sun et~al\mbox{.}(2023)]%
        {sun2023automatic}
\bibfield{author}{\bibinfo{person}{Weisong Sun}, \bibinfo{person}{Chunrong Fang}, \bibinfo{person}{Yudu You}, \bibinfo{person}{Yun Miao}, \bibinfo{person}{Yi Liu}, \bibinfo{person}{Yuekang Li}, \bibinfo{person}{Gelei Deng}, \bibinfo{person}{Shenghan Huang}, \bibinfo{person}{Yuchen Chen}, \bibinfo{person}{Quanjun Zhang}, {et~al\mbox{.}}} \bibinfo{year}{2023}\natexlab{}.
\newblock \showarticletitle{Automatic code summarization via chatgpt: How far are we?}
\newblock \bibinfo{journal}{\emph{arXiv preprint arXiv:2305.12865}} (\bibinfo{year}{2023}).
\newblock


\bibitem[Tibshirani and Efron(1993)]%
        {tibshirani1993introduction}
\bibfield{author}{\bibinfo{person}{Robert~J Tibshirani} {and} \bibinfo{person}{Bradley Efron}.} \bibinfo{year}{1993}\natexlab{}.
\newblock \showarticletitle{An introduction to the bootstrap}.
\newblock \bibinfo{journal}{\emph{Monographs on statistics and applied probability}} \bibinfo{volume}{57}, \bibinfo{number}{1} (\bibinfo{year}{1993}), \bibinfo{pages}{1--436}.
\newblock


\bibitem[Tosun et~al\mbox{.}(2009)]%
        {tosun2009practical}
\bibfield{author}{\bibinfo{person}{Ay{\c{s}}e Tosun}, \bibinfo{person}{Burak Turhan}, {and} \bibinfo{person}{Ay{\c{s}}e Bener}.} \bibinfo{year}{2009}\natexlab{}.
\newblock \showarticletitle{Practical considerations in deploying ai for defect prediction: a case study within the turkish telecommunication industry}. In \bibinfo{booktitle}{\emph{Proceedings of the 5th International Conference on Predictor Models in Software Engineering}}. \bibinfo{pages}{1--9}.
\newblock


\bibitem[Wahono(2015)]%
        {wahono2015systematic}
\bibfield{author}{\bibinfo{person}{Romi~Satria Wahono}.} \bibinfo{year}{2015}\natexlab{}.
\newblock \showarticletitle{A systematic literature review of software defect prediction}.
\newblock \bibinfo{journal}{\emph{Journal of software engineering}} \bibinfo{volume}{1}, \bibinfo{number}{1} (\bibinfo{year}{2015}), \bibinfo{pages}{1--16}.
\newblock


\bibitem[Wang et~al\mbox{.}(2012)]%
        {wang2012compressed}
\bibfield{author}{\bibinfo{person}{Jun Wang}, \bibinfo{person}{Beijun Shen}, {and} \bibinfo{person}{Yuting Chen}.} \bibinfo{year}{2012}\natexlab{}.
\newblock \showarticletitle{Compressed C4. 5 models for software defect prediction}. In \bibinfo{booktitle}{\emph{2012 12th International Conference on quality software}}. IEEE, \bibinfo{pages}{13--16}.
\newblock


\bibitem[Wang et~al\mbox{.}(2018)]%
        {wang2018deep}
\bibfield{author}{\bibinfo{person}{Song Wang}, \bibinfo{person}{Taiyue Liu}, \bibinfo{person}{Jaechang Nam}, {and} \bibinfo{person}{Lin Tan}.} \bibinfo{year}{2018}\natexlab{}.
\newblock \showarticletitle{Deep semantic feature learning for software defect prediction}.
\newblock \bibinfo{journal}{\emph{IEEE Transactions on Software Engineering}} \bibinfo{volume}{46}, \bibinfo{number}{12} (\bibinfo{year}{2018}), \bibinfo{pages}{1267--1293}.
\newblock


\bibitem[Wang et~al\mbox{.}(2023)]%
        {wang2023codet5+}
\bibfield{author}{\bibinfo{person}{Yue Wang}, \bibinfo{person}{Hung Le}, \bibinfo{person}{Akhilesh~Deepak Gotmare}, \bibinfo{person}{Nghi~DQ Bui}, \bibinfo{person}{Junnan Li}, {and} \bibinfo{person}{Steven~CH Hoi}.} \bibinfo{year}{2023}\natexlab{}.
\newblock \showarticletitle{Codet5+: Open code large language models for code understanding and generation}.
\newblock \bibinfo{journal}{\emph{arXiv preprint arXiv:2305.07922}} (\bibinfo{year}{2023}).
\newblock


\bibitem[Wang et~al\mbox{.}(2024)]%
        {wang2024line}
\bibfield{author}{\bibinfo{person}{Ziliang Wang}, \bibinfo{person}{Ge Li}, \bibinfo{person}{Jia Li}, \bibinfo{person}{Yihong Dong}, \bibinfo{person}{Yingfei Xiong}, {and} \bibinfo{person}{Zhi Jin}.} \bibinfo{year}{2024}\natexlab{}.
\newblock \showarticletitle{Line-level Semantic Structure Learning for Code Vulnerability Detection}.
\newblock \bibinfo{journal}{\emph{arXiv preprint arXiv:2407.18877}} (\bibinfo{year}{2024}).
\newblock


\bibitem[Wilcoxon(1945)]%
        {wilcoxon1945individual}
\bibfield{author}{\bibinfo{person}{Frank Wilcoxon}.} \bibinfo{year}{1945}\natexlab{}.
\newblock \showarticletitle{Individual comparisons by ranking methods}.
\newblock \bibinfo{journal}{\emph{Biometrics bulletin}} \bibinfo{volume}{1}, \bibinfo{number}{6} (\bibinfo{year}{1945}), \bibinfo{pages}{80--83}.
\newblock


\bibitem[Xia et~al\mbox{.}(2023)]%
        {xia2023automated}
\bibfield{author}{\bibinfo{person}{Chunqiu~Steven Xia}, \bibinfo{person}{Yuxiang Wei}, {and} \bibinfo{person}{Lingming Zhang}.} \bibinfo{year}{2023}\natexlab{}.
\newblock \showarticletitle{Automated program repair in the era of large pre-trained language models}. In \bibinfo{booktitle}{\emph{2023 IEEE/ACM 45th International Conference on Software Engineering (ICSE)}}. IEEE, \bibinfo{pages}{1482--1494}.
\newblock


\bibitem[Xiao et~al\mbox{.}(2023)]%
        {xiao2023empirical}
\bibfield{author}{\bibinfo{person}{Yan Xiao}, \bibinfo{person}{Xinyue Zuo}, \bibinfo{person}{Lei Xue}, \bibinfo{person}{Kailong Wang}, \bibinfo{person}{Jin~Song Dong}, {and} \bibinfo{person}{Ivan Beschastnikh}.} \bibinfo{year}{2023}\natexlab{}.
\newblock \showarticletitle{Empirical study on transformer-based techniques for software engineering}.
\newblock \bibinfo{journal}{\emph{arXiv preprint arXiv:2310.00399}} (\bibinfo{year}{2023}).
\newblock


\bibitem[Yan et~al\mbox{.}(2019)]%
        {yan2019characterizing}
\bibfield{author}{\bibinfo{person}{Meng Yan}, \bibinfo{person}{Xin Xia}, \bibinfo{person}{David Lo}, \bibinfo{person}{Ahmed~E Hassan}, {and} \bibinfo{person}{Shanping Li}.} \bibinfo{year}{2019}\natexlab{}.
\newblock \showarticletitle{Characterizing and identifying reverted commits}.
\newblock \bibinfo{journal}{\emph{Empirical Software Engineering}} \bibinfo{volume}{24}, \bibinfo{number}{4} (\bibinfo{year}{2019}), \bibinfo{pages}{2171--2208}.
\newblock


\bibitem[Yang et~al\mbox{.}(2025)]%
        {yang2025qwen3}
\bibfield{author}{\bibinfo{person}{An Yang}, \bibinfo{person}{Anfeng Li}, \bibinfo{person}{Baosong Yang}, \bibinfo{person}{Beichen Zhang}, \bibinfo{person}{Binyuan Hui}, \bibinfo{person}{Bo Zheng}, \bibinfo{person}{Bowen Yu}, \bibinfo{person}{Chang Gao}, \bibinfo{person}{Chengen Huang}, \bibinfo{person}{Chenxu Lv}, {et~al\mbox{.}}} \bibinfo{year}{2025}\natexlab{}.
\newblock \showarticletitle{Qwen3 technical report}.
\newblock \bibinfo{journal}{\emph{arXiv preprint arXiv:2505.09388}} (\bibinfo{year}{2025}).
\newblock


\bibitem[Yang et~al\mbox{.}(2015)]%
        {yang2015deep}
\bibfield{author}{\bibinfo{person}{Xinli Yang}, \bibinfo{person}{David Lo}, \bibinfo{person}{Xin Xia}, \bibinfo{person}{Yun Zhang}, {and} \bibinfo{person}{Jianling Sun}.} \bibinfo{year}{2015}\natexlab{}.
\newblock \showarticletitle{Deep learning for just-in-time defect prediction}. In \bibinfo{booktitle}{\emph{2015 IEEE International conference on software quality, reliability and security}}. IEEE, \bibinfo{pages}{17--26}.
\newblock


\bibitem[Ye et~al\mbox{.}(2023)]%
        {ye2023tram}
\bibfield{author}{\bibinfo{person}{Tong Ye}, \bibinfo{person}{Lingfei Wu}, \bibinfo{person}{Tengfei Ma}, \bibinfo{person}{Xuhong Zhang}, \bibinfo{person}{Yangkai Du}, \bibinfo{person}{Peiyu Liu}, \bibinfo{person}{Shouling Ji}, {and} \bibinfo{person}{Wenhai Wang}.} \bibinfo{year}{2023}\natexlab{}.
\newblock \showarticletitle{Tram: A token-level retrieval-augmented mechanism for source code summarization}.
\newblock \bibinfo{journal}{\emph{arXiv preprint arXiv:2305.11074}} (\bibinfo{year}{2023}).
\newblock


\bibitem[Zhang et~al\mbox{.}(2025a)]%
        {zhang2025revisiting}
\bibfield{author}{\bibinfo{person}{Ting Zhang}, \bibinfo{person}{Ivana~Clairine Irsan}, \bibinfo{person}{Ferdian Thung}, {and} \bibinfo{person}{David Lo}.} \bibinfo{year}{2025}\natexlab{a}.
\newblock \showarticletitle{Revisiting sentiment analysis for software engineering in the era of large language models}.
\newblock \bibinfo{journal}{\emph{ACM Transactions on Software Engineering and Methodology}} \bibinfo{volume}{34}, \bibinfo{number}{3} (\bibinfo{year}{2025}), \bibinfo{pages}{1--30}.
\newblock


\bibitem[Zhang et~al\mbox{.}(2025b)]%
        {zhang2025benchmarking}
\bibfield{author}{\bibinfo{person}{Ting Zhang}, \bibinfo{person}{Chengran Yang}, \bibinfo{person}{Yindu Su}, \bibinfo{person}{Martin Weyssow}, \bibinfo{person}{Hung Nguyen}, \bibinfo{person}{Tan Bui}, \bibinfo{person}{Hong~Jin Kang}, \bibinfo{person}{Yikun Li}, \bibinfo{person}{Eng~Lieh Ouh}, \bibinfo{person}{Lwin~Khin Shar}, {et~al\mbox{.}}} \bibinfo{year}{2025}\natexlab{b}.
\newblock \showarticletitle{Benchmarking Large Language Models for Multi-Language Software Vulnerability Detection}.
\newblock \bibinfo{journal}{\emph{arXiv preprint arXiv:2503.01449}} (\bibinfo{year}{2025}).
\newblock


\bibitem[Zhou et~al\mbox{.}(2019)]%
        {zhou2019devign}
\bibfield{author}{\bibinfo{person}{Yaqin Zhou}, \bibinfo{person}{Shangqing Liu}, \bibinfo{person}{Jingkai Siow}, \bibinfo{person}{Xiaoning Du}, {and} \bibinfo{person}{Yang Liu}.} \bibinfo{year}{2019}\natexlab{}.
\newblock \showarticletitle{Devign: Effective vulnerability identification by learning comprehensive program semantics via graph neural networks}.
\newblock \bibinfo{journal}{\emph{Advances in neural information processing systems}}  \bibinfo{volume}{32} (\bibinfo{year}{2019}).
\newblock


\end{thebibliography}
\end{document}